\documentclass[10pt]{article}
\usepackage[subpreambles=false]{standalone}
\usepackage{import}
\usepackage[utf8]{inputenc}
\usepackage{url}
\usepackage{graphicx}
\usepackage{multirow}
\usepackage[colorinlistoftodos]{todonotes}
\usepackage{xspace} 
\usepackage{framed} 
\usepackage{mdframed}
\usepackage{subfigure}
\usepackage{cancel}
\usepackage{siunitx}
\usepackage{efbox,graphicx}
\usepackage[most]{tcolorbox}
\usepackage{amsmath}
\usepackage{mathtools,amssymb}
\usepackage{multirow}
\usepackage{makecell}
\usepackage{enumitem}
\definecolor{highlight}{RGB}{255, 138, 171}
\efboxsetup{linecolor=highlight,linewidth=1pt}
\usepackage{lineno,hyperref}
\usepackage{parcolumns}
\definecolor{listinggray}{gray}{0.9}
\definecolor{lbcolor}{rgb}{1,1,1}
\usepackage{wrapfig}

\lstdefinestyle{customc}{
  backgroundcolor=\color{black!5},
  belowcaptionskip=1\baselineskip,
  breaklines=true,
  frame=L,
  xleftmargin=0.2cm,
  language=C,
  showstringspaces=false,
  basicstyle=\footnotesize\ttfamily,
  keywordstyle=\bfseries\color{green!40!black},
  commentstyle=\itshape\color{purple!40!black},
  identifierstyle=\color{blue},
  stringstyle=\color{orange},
}

\lstdefinestyle{customasm}{
  belowcaptionskip=1\baselineskip,
  frame=L,
  language=[x86masm]Assembler,
  basicstyle=\footnotesize\ttfamily,
  commentstyle=\itshape\color{purple!40!black},
}

\definecolor{jonas}{RGB}{128, 1, 0}
\definecolor{ForestGreen}{RGB}{63,142,38}

\newcommand{\AM}[1]{{\color{black}#1}}

\newcommand{\rwone}[1]{{\color{black}#1}}

\newcommand{\rwthree}[1]{{\color{black}#1}}

\newcommand{\cut}[1]{}

\definecolor{color_greenish}{rgb}{0.0, 0.5, 0.0}
\definecolor{darkgreen}{RGB}{17, 89, 9}
\definecolor{color_fmc}{RGB}{245, 14, 78} 

\definecolor{color_mb}{rgb}{0.9, 0.3, 0.0} 

\newcommand{\openmp}{Open{MP}\xspace}
\newcommand{\lbomp}{LB4OMP\xspace}
\newcommand{\gromacs}{\mbox{GROMACS}\xspace}
\newcommand{\sphynx}{\mbox{SPHYNX}\xspace}

\newcommand{\dist}{\mbox{DIST}\xspace}
\newcommand{\stream}{\mbox{STREAM}\xspace}
\newcommand{\xeon}{miniHPC-Broadwell\xspace}
\newcommand{\knl}{miniHPC-KNL\xspace}
\newcommand{\daint}{Piz Daint-Haswell\xspace}
\newcommand{\cov}{\emph{c.o.v.}\xspace}
\newcommand{\percentimbalance}{\emph{p.i.}\xspace}
\newcommand{\awfb}{\texttt{\mbox{AWF-B}}\xspace}
\newcommand{\awfc}{\texttt{\mbox{AWF-C}}\xspace}
\newcommand{\awfd}{\texttt{\mbox{AWF-D}}\xspace}
\newcommand{\awfe}{\texttt{\mbox{AWF-E}}\xspace}

\usepackage{verbatim}

\newcommand{\timenoloop}{$T\_ol$\xspace}
\definecolor{mypink3}{cmyk}{0, 0.7808, 0.4429, 0.1412}

\hypersetup{
    colorlinks,
    linkcolor={black},
    citecolor={black},
    urlcolor={blue!80!black}
}

  \usepackage{cite}

\begin{document}
\title{LB4OMP: A Dynamic Load Balancing Library for \\ Multithreaded Applications}

\author{Jonas H. M\"uller Kornd\"orfer$^{\ast}$,
        Ahmed Eleliemy$^{\ast}$,\\
        Ali Mohammed$^{\ast}$$^{\dagger}$,
        Florina M. Ciorba$^{\ast}$\\
        $^{\ast}$ Department of Mathematics and Computer Science, \\University of Basel
        , Switzerland\\
        $^{\dagger}$ HPE’s HPC/AI EMEA Research Lab (ERL), Switzerland\\
        Email: $^{\ast}$\{firstname.lastname\}@unibas.ch \\ $^{\dagger}$ali.mohammed@hpe.com
        }



\maketitle
\sloppy
\begin{abstract}
	
	Exascale computing systems will exhibit high degrees of hierarchical parallelism, with thousands of computing nodes and hundreds of cores per node.
	Efficiently exploiting hierarchical parallelism is challenging due to load imbalance that arises at multiple levels. 
	OpenMP is the most widely-used standard for expressing and exploiting the ever-increasing node-level parallelism. 
	The scheduling options in OpenMP are insufficient to address the load imbalance that arises during the execution of multithreaded applications. 
	The limited scheduling options in OpenMP hinder research on novel scheduling techniques which require comparison with others from the literature. 
	This work introduces LB4OMP, an open-source dynamic load balancing library that implements successful scheduling algorithms from the literature. 
	LB4OMP is a research infrastructure designed to spur and support present and future scheduling research, for the benefit of multithreaded applications performance. 
	Through an extensive performance analysis campaign, we assess the effectiveness and demystify the performance of all loop scheduling techniques in the library. 
	We show that, for numerous applications-systems pairs, the scheduling techniques in LB4OMP outperform the scheduling options in OpenMP. 
	Node-level load balancing using LB4OMP leads to reduced cross-node load imbalance and to improved MPI+OpenMP applications performance, which is critical for Exascale computing.
	
\end{abstract}

\setcounter{figure}{0}
\section{Introduction} \label{sec:intro}

On the road to Exascale, we observe that modern and future high performance computing~(HPC) systems combine an increasing number of computing nodes and, in particular, cores per node.
For example, the top 5 systems on the Top500 list\footnote{\href{https://www.top500.org/lists/top500/2020/11/}{www.top500.org/lists/top500/2020/11/}} contain thousands of nodes and tens to hundreds of cores per node.

Recent reports\footnote{\href{https://www.nextplatform.com/2021/02/10/a-sneak-peek-at-chinas-sunway-exascale-supercomputer/}{www.nextplatform.com/2021/02/10/a-sneak-peek-at-chinas-sunway-exascale-supercomputer/}} indicate that the next update of the Sunway TaihuLight system will include $520$ cores per node, double that of its predecessor. 
Such hardware parallelism increase leads to the challenge of exposing and expressing corresponding degrees of hierarchical parallelism in software to efficiently exploit the hierarchical hardware parallelism. 


Load imbalance is 
a significant performance degradation factor in \mbox{\textit{computationally-intensive}} applications~\cite{hummel1996load}\cite{dongarra2011international}, defined as processors idling while there exist units of computation ready to be executed that no processor has started. 
This results in uneven execution progress among the parallel processing units, which can emerge from numerous application-, algorithm-, and/or systemic characteristics.
Computationally-intensive applications often represent \textit{irregular} workloads (e.g., due to boundary conditions, convergence, conditions, and branches). 
Computing systems may consist of \textit{heterogeneous} processors and may be affected by nonuniform memory access (NUMA) times, operating system noise, and contention due to sharing of resources.
Load imbalance can be mitigated by an efficient dynamic scheduling of computation units onto processing units. 
Finding optimal schedules is \mbox{NP-hard}~\cite{johnson1985np}.
Therefore, many scheduling heuristics have been proposed over the years\cite{khan1994comparison}\cite{izakian2009comparison}.

Scheduling and load balancing that exploit multiple levels of hardware parallelism across and within computing nodes 
are critical challenges for the upcoming Exascale systems~\cite{bergman2008exascale}\cite{asch2018big}.
Dynamic \mbox{self-scheduling} explicitly addresses application- and \mbox{system-induced} performance variations \AM{while minimizing load imbalance and} scheduling overhead~\cite{AF:2000}\cite{Automatic-OMP-LS:2012}\cite{Knowledge-BasedAdaptive2012}.

It has been recently shown that thread-level load imbalance has a significant impact on the performance of hybrid MPI+OpenMP applications~\cite{ali2020twoLevel}.
\openmp is the most \mbox{widely-used} standard for expressing and exploiting node-level parallelism.
The \openmp standard specifies three loop schedule kinds: \texttt{static}, \texttt{dynamic}, and \texttt{guided}.
These scheduling options limit the highest achievable performance as they do not cover the broad spectrum of applications and systems characteristics~\cite{Ayguade2003}~\cite{ciorba:2018}\cite{Kasielke:2019}. 
Furthermore, the absence of a comparative implementation of the multitude of scheduling techniques from the literature hinders research on novel scheduling techniques which typically requires comparison with the scheduling state of the art.

This work builds on recent work on multilevel load balancing~\cite{ali2020twoLevel} by concentrating on thread-level scheduling and deepening the analysis of its performance impact for multithreaded applications executing on hierarchical parallel systems. 
Specifically, this work \textbf{provides} a broad range of dynamic loop self-scheduling~(DLS) techniques, implemented in a unified OpenMP runtime library (RTL), called \lbomp\footnote{\href{https://github.com/unibas-dmi-hpc/LB4OMP}{github.com/unibas-dmi-hpc/LB4OMP}} that can readily be used for MPI+\openmp applications.
Aiming for a \textbf{wide reach} and \textbf{broad impact}, we implemented \lbomp as an extension of LLVM's \openmp RTL given its widespread use (e.g., in the US DOE's Exascale Computing Project project~\cite{heroux2020ecp}), \mbox{open-source} nature, and high compatibility with widely-used compilers (Intel, IBM, PGI, GNU).
%

The \lbomp library supports 14 carefully selected \textit{dynamic} (and \textit{adaptive}) loop self-scheduling techniques, ready to use in addition to those existing in the standard-compliant \openmp libraries.
Applications using \lbomp benefit from improved performance due to the portfolio of DLS techniques, of which certain \textit{adapt}\textbf{} during execution to unpredictable variations in application and systemic characteristics (see Section~\ref{subsec:DLS}).
These 14 techniques are selected to cover a broad spectrum of \textit{dynamic} (and \textit{adaptive}) scheduling techniques. Specifically, \lbomp provides:
\begin{enumerate}[align=left, leftmargin=*]
    \item[$\bullet$] Five \textit{dynamic} but \textbf{\textit{\mbox{non-adaptive}}} \mbox{self-scheduling} techniques: fixed size chunking~(\texttt{FSC})~\cite{FSC:1985}, factoring~(\texttt{FAC})~\cite{FAC:1992}, the practical variant of factoring~(\texttt{FAC2}), tapering~(\texttt{TAP})~\cite{TAP:1992}, and the practical variant of weighted factoring~(\texttt{WF2})~\cite{WF:1996};
    \item[$\bullet$] Seven \textit{dynamic} and \textbf{\mbox{\textit{adaptive}}} \mbox{self-scheduling} techniques: \texttt{BOLD}~\cite{BOLD:1997}, adaptive weighted factoring~(\texttt{AWF}), its variants (\texttt{AWF-}\texttt{B},\texttt{C},\texttt{D},\texttt{E})~\cite{AWF:2003}, and adaptive factoring~(\texttt{AF})~\cite{AF:2000};
    \item[$\bullet$] Features for performance measurement to measure loop-specific performance metrics for an in-depth analysis of loop scheduling and load balancing.
\end{enumerate}
\rwone{\lbomp also provides \texttt{mFAC} and \texttt{mAF}, two improved implementations that reduce the overhead of \texttt{FAC} and \texttt{AF}.}

The scheduling techniques in \lbomp differ in the amount of work assigned to a thread at a time, referred to as a chunk of loop iterations. 
Specifically techniques with: (1)~\textit{simple} chunk calculation include \texttt{FSC}, \texttt{FAC2}, and \texttt{WF2};
(2)~\textit{\mbox{profiling-based}} chunk calculation include \texttt{FAC}, \texttt{mFAC}, and \texttt{TAP}; 
and (3)~\textit{adaptive} (nonlinear) chunk calculation include \texttt{BOLD}, \texttt{AWF}, its variants \texttt{AWF-}\texttt{B},\texttt{C},\texttt{D},\texttt{E}, and \texttt{AF} and \texttt{mAF}. 
\noindent This work makes the following \textbf{contributions}:
\begin{enumerate}[align=left, leftmargin=*]
\item \emph{A novel systematic and unified implementation} of 14 \textit{dynamic} (and \textit{adaptive}) scheduling techniques. 
\item \emph{Advanced features for performance measurement} of loop performance and loop-level load imbalance. 
\item \emph{An in-depth analysis of the performance potential and limitations} of the standard and newly implemented scheduling techniques, which were so far only partially known to the non-experts and/or scheduling practitioners.
\end{enumerate}
%
\rwone{
The \textbf{novelty} of this work lies in providing a standalone and unified implementation of efficient scheduling techniques from literature, which is needed to \textbf{spur new research in scheduling and load balancing for Exascale systems}. 
Prior to this work, scheduling research was hindered by the absence of an environment that supports a fair comparison with the existing scheduling algorithms in the literature.
Novel scheduling techniques or improved versions of the techniques implemented in this work can now be implemented and compared in this unified testbed.
LB4OMP enables and promotes research on automatic selection methods to identify, during execution, the highest performing scheduling technique for a given application-loop-\textit{time-step} configuration.


This work is \textbf{significant} by \emph{bridging the gap} between the \mbox{state-of-the-art} and the \mbox{state-of-the-practice} of load balancing in multithreaded applications.
This will allow the large degrees of heterogeneous node-level parallelism in today's pre- and upcoming Exascale systems to be efficiently exploited for improving applications performance. 
}

This work is organized as follows. 
Section~\ref{sec:rw} is a review of the related literature highlighting the differences between prior and the present work.
Section~\ref{sec:dlsLLVM} describes the \lbomp~design and highlights the required extensions to LLVM's \openmp RTL. 
The use of the newly implemented scheduling techniques in \openmp applications via the \lbomp~library is detailed in Section~\ref{sec:dlsLLVM}.
The experimental design and performance analysis campaign are presented and discussed in Section~\ref{sec:experiments}. 
The work is concluded in Section~\ref{sec:conclusion}, which also outlines directions for future work.


\section{Related Work}\label{rw}
\label{sec:rw}
The performance potential of a small number of \textit{dynamic} and \textbf{\textit{non-adaptive}} loop scheduling techniques (\texttt{TSS}, \texttt{FAC2}, \texttt{WF2}, and \texttt{RAND}), implemented in the GNU \openmp RTL, was recently explored~\cite{ciorba:2018}. 
The authors showed cases when applications achieve improved performance beyond the one offered by the scheduling techniques supported in the GNU \openmp RTL. 
Another variant of \texttt{FAC}, called \texttt{BO FSS}, was proposed and compared against \texttt{STATIC}, \texttt{GSS}, \texttt{TSS}, \texttt{FAC2}, \texttt{TAP}~\cite{TAP:1992}, \texttt{HSS}~\cite{HSS2006}, and \texttt{BinLPT}~\cite{penna2017binlpt}, in another extended implementation in the GNU \openmp RTL~\cite{facProb2020}. 
The scheduling techniques considered in these research efforts does not consider \textit{dynamic and \textbf{adaptive}} scheduling techniques. 
In general, other efforts only considered extending the GNU \openmp RTL, which is not compatible with other compilers, unlike the LLVM \openmp RTL.

LLVM has gained considerable traction in the software vendor community and improving the open source LLVM compiler and runtime ecosystem is a priority for the US DOE Exascale Computing Project~\cite{heroux2020ecp}.
The LLVM \openmp RTL was extended only by an implementation of \texttt{FAC2}~\cite{Kasielke:2019}.
Experiments therein showed improved performance of certain workloads with the newly added DLS technique.

The present work improves over the previous related work by 
(1)~providing \emph{a novel systematic and unified implementation} of a broader range of \textit{dynamic} (and \textit{adaptive}) scheduling techniques.
(2)~providing \emph{advanced features for performance measurement} of loop performance and loop-level load imbalance.
(3)~\emph{demystifying the performance potential and limitations} of the standard and the newly implemented scheduling techniques through \emph{an in-depth performance analysis campaign}. 


Another direction of related work includes efforts that propose generic interfaces to allow users to implement their own loop scheduling techniques in different runtime libraries~\cite{SeonmyeongOPTexecParUDS}~\cite{kale2019toward}~\cite{santana:hal-02454426}.
These efforts reduce the development challenges associated with the direct modification to the RTL source codes, i.e., developers can implement their scheduling technique via simplified, and ideally, \mbox{well-documented} interfaces.
However, these efforts do not exclude the need for extensive scheduling libraries to validate novel scheduling techniques and exploit the increased hardware parallelism of modern HPC systems.
Therefore, such efforts~\cite{SeonmyeongOPTexecParUDS,kale2019toward,santana:hal-02454426} can be seen as potential methods that facilitate the development of another version of the \lbomp scheduling library in the future.

Considering the vast amount of DLS techniques proposed in the literature, the following non-trivial question arises: 
\textit{What are the criteria to include a particular scheduling technique into a unified scheduling library?}
A number of research efforts attempted to answer this question~\cite{TSS:1993}~\cite{FAC:1992}~\cite{TAP:1992}~\cite{BOLD:1997}~\cite{AF:2000}~\cite{AWF:2003}.

\texttt{FSC}~\cite{FSC-DataLoc:1996}, \texttt{TSS}~\cite{TSS:1993}, \texttt{FAC}~\cite{FAC:1992}, and \texttt{TAP}~\cite{TAP:1992} were introduced by separate research groups.
However, the performance of each of these techniques was compared against at least one of three main scheduling techniques \texttt{STATIC}, \texttt{SS}~\cite{SS}, and \texttt{GSS}~\cite{GSS:1987} that nowadays correspond to \texttt{schedule(static)}, \texttt{schedule(dynamic,1)}, and \texttt{schedule(guided,1)} specified in the \openmp standard.
The chunk calculation in \texttt{TSS}, for instance, is simpler than the \mbox{non-linear} chunk calculation in \texttt{GSS}. 
Therefore, we state that the \textit{simplicity of chunk calculation} is an important selection criterion.

\texttt{FSC}, \texttt{TAP}, and \texttt{FAC} are based on probabilistic analyses and use profiling information to calculate the chunk sizes that achieve the most balanced load execution for a given application with a high probability.
The profiling information is obtained prior to the applications' execution. 
We state that the \textit{\mbox{profiling-based} chunk calculation} is another important selection criterion.


Another set of related research efforts includes \texttt{BOLD}~\cite{BOLD:1997}, \texttt{AWF-}\texttt{B},\texttt{C},\texttt{D},\texttt{E}~\cite{AWF:2003}, and \texttt{AF}~\cite{AF:2000}. 
These techniques use profiling information obtained during applications' execution to adapt during execution the calculated chunk sizes to achieve the most balanced load execution for a given application.
We state that the \textit{adaptivity of chunk calculation} is another valuable selection criterion.

Based on the above criteria, \texttt{FSC}~\cite{FSC-DataLoc:1996}, \texttt{FAC}~\cite{FAC:1992}, \texttt{TAP}~\cite{TAP:1992}, \texttt{WF}~\cite{WF:1996}, \texttt{BOLD}~\cite{BOLD:1997}, \texttt{AWF-}\texttt{B},\texttt{C},\texttt{D},\texttt{E}~\cite{AWF:2003}, and \texttt{AF}~\cite{AF:2000} were selected for implementation into the \lbomp scheduling library.
Other scheduling techniques that meet these criteria can also be considered for inclusion in \lbomp.
The DLS techniques selected in this work can also be applied to schedule OpenMP \texttt{tasks} and \texttt{taskloops}~\cite{duran2008evaluation}\cite{clet2014evaluation}. 
The use of LB4OMP for scheduling \texttt{tasks} and \texttt{taskloops} is part of separate ongoing work by the authors.


\section{The \lbomp Library}\label{sec:dlsLLVM}


\lbomp extends the LLVM \openmp RTL version 8.0, which is widely used and compatible with various compilers, including Intel, IBM, GNU, and PGI.
The choice of extending the LLVM \openmp RTL meets important research goals and priorities of the HPC community~\cite{heroux2020ecp}.

\begin{figure}[!htb]
		\centering
		\includegraphics[width=0.85\columnwidth]{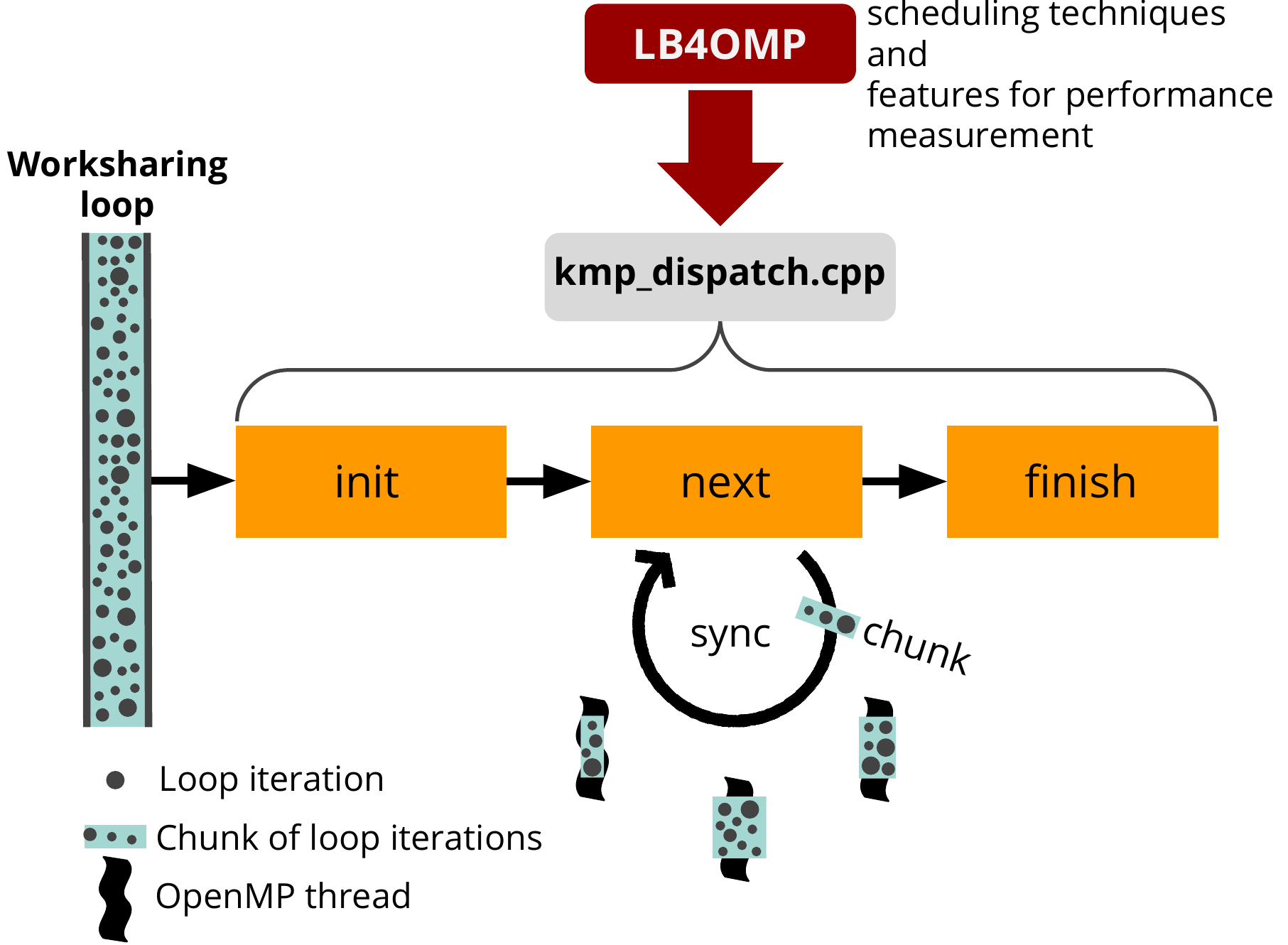} 
		\caption[]{Extension of the \openmp LLVM RTL scheduling process for worksharing loops with \lbomp.
		}
		\label{fig:dlsLLVM}
		\vspace{-0.1cm}
	\end{figure}

\figurename~\ref{fig:dlsLLVM} shows the \lbomp loop scheduling mechanism which extends the scheduling mechanism in the LLVM \openmp RTL. 
The three main functions responsible for the chunk calculation are implemented in the file \texttt{kmp\_dispatch.cpp}. 
Upon initialization, each thread calls the \texttt{\_\_kmp\_dispatch\_init\_algorithm} function inside the \texttt{kmp\_dispatch.cpp} file (init in \figurename~\ref{fig:dlsLLVM}). 
This function then initializes the needed structures for the selected scheduling technique and calls \texttt{\_\_kmp\_dispatch\_next\_algorithm} (next in \figurename~\ref{fig:dlsLLVM}).
The logic of the chunk calculation of all DLS techniques is implemented in the \texttt{\_\_kmp\_dispatch\_next\_algorithm} function. 
The \texttt{\_\_kmp\_dispatch\_next\_algorithm} is called each time a thread needs to obtain work.
Since the threads obtain work from a shared queue, \texttt{\_\_kmp\_dispatch\_next\_algorithm} relies on different synchronization operations (sync in \figurename~\ref{fig:dlsLLVM}) depending on the scheduling technique in execution.
Finally, the threads call \texttt{\_\_kmp\_dispatch\_finish} (finish in \figurename~\ref{fig:dlsLLVM}) to reset variables or free allocated memory. 
	
\noindent\textbf{Significance of chunk parameter.}
The \openmp standard scheduling techniques and the newly implemented scheduling techniques in \lbomp support the declaration of a \textit{chunk parameter} which bears different meanings among the scheduling techniques. 
For \mbox{\texttt{schedule(static, chunk)}} and \mbox{\texttt{schedule(dynamic,chunk)}}, the chunk parameter denotes the amount of iterations that the threads should receive for every work request.
For the other techniques, the chunk parameter works as a \textbf{threshold}, in the sense that when chunks sizes, calculated by a scheduling technique, fall below this threshold they will be replaced by a chunk sizes equal to the size of the chunk parameter. 
The chunk parameter was introduced by the \openmp standard to minimize the scheduling overhead and to improve data locality. 
Declaring a proper \textit{chunk parameter} improves performance since threads perform fewer scheduling rounds than without this threshold.
This is confirmed by the experiments described in Section~\ref{sec:experiments}.

%

\subsection{Dynamic Loop Scheduling Techniques}
\label{subsec:DLS}
	\lbomp \emph{bridges the gap} between the literature and the practice in dynamic load balancing of multithreaded applications. 
	It represents an environment for a fair comparison of scheduling techniques and lays the ground for future research in loop scheduling of multithreaded applications.	

	The loop scheduling techniques implemented in \lbomp are \emph{dynamic} and a number of them are also \emph{adaptive} \mbox{self-scheduling} techniques. 
	With \mbox{self-scheduling} techniques, free threads request, calculate, and obtain their own next chunk of units of work (loop iterations) by accessing a central shared work queue containing all iterations of a given loop. 
	The chunk size is calculated according to the loop scheduling techniques in the \openmp RTL.


Following is a brief description of each scheduling technique in \lbomp, starting with \textit{dynamic} and \textit{\mbox{\textbf{non-adaptive}}} \mbox{self-scheduling} techniques followed by the \textit{dynamic} and \textit{\textbf{adaptive}} \mbox{self-scheduling} techniques.
More details about the various chunk calculations for these techniques can be found in the literature~\cite{eleliemy2021distributed}.
\noindent \textbf{Dynamic and non-adaptive self-scheduling.}\\
\texttt{SS} (or \texttt{dynamic,1} in \openmp)~\cite{SS} is a dynamic \mbox{self-scheduling} technique wherein the chunk size is always one loop iteration. 
\texttt{SS} incurs the highest scheduling overhead due to the largest number of chunks (equal to the number of loop iterations). 
\texttt{SS} can achieve a highly \mbox{load-balanced} execution in highly irregular execution environments.

\texttt{FSC}~\cite{FSC:1985} determines an optimal chunk size that achieves a balanced execution of loop iterations with the smallest overhead.
To calculate the optimal chunk size, \texttt{FSC} requires that the variability in iteration execution times and the scheduling overhead of assigning loop iterations are known before applications' execution.

\texttt{GSS}~\cite{GSS:1987} is a trades-off between the load balancing achievable with \texttt{SS} and the low scheduling overhead incurred by \texttt{STATIC}.
Unlike \texttt{FSC}, \texttt{GSS} assigns decreasing chunk sizes to balance the loop execution progress among all threads.
For every work request, \texttt{GSS} assigns a chunk size equal to the number of remaining loop iterations divided by the total number of threads.

\texttt{TAP}~\cite{TAP:1992} is based on a probabilistic analysis that represents a general case of \texttt{GSS}.
It considers the average of loop iteration execution times~$\mu$ and the standard deviation~$\sigma$ to achieve a higher load balance than \texttt{GSS}. 

\texttt{TSS}~\cite{TSS:1993} assigns decreasing chunk sizes similar to \texttt{GSS}. 
However, \texttt{TSS} uses a linear function to decrement chunk sizes. 
This linearity results in lower scheduling overhead in each scheduling step compared to \texttt{GSS}.

\texttt{FAC}~\cite{FAC:1992} schedules the loop iterations in batches of equally-sized chunks. 
\texttt{FAC} evolved from comprehensive probabilistic analyses, and assumes prior knowledge about the average iteration execution times ($\mu$) and their standard deviation ($\sigma$). 
A practical implementation of \texttt{FAC}, denoted \texttt{FAC2}, assigns half of the remaining loop iterations for every batch. 
The initial chunk size of \texttt{FAC2} is half of the initial chunk size of \texttt{GSS}. 
If more \mbox{time-consuming} loop iterations are at the beginning of the loop, \texttt{FAC2} is expected to better balance their execution than \texttt{GSS}.

\texttt{WF}~\cite{WF:1996} is similar to \texttt{FAC}, with the difference that each processing unit executes variably-sized chunks of a given batch according to its relative processing weights. 
The processing weights, $W_{pj}$, are determined prior to applications’ execution and remain constant during execution.
\texttt{WF2} is the practical implementation of \texttt{WF} that is based on \texttt{FAC2}.

\texttt{mFAC} is our improved implementation of \texttt{FAC}. In the original \texttt{FAC} algorithm, the first thread that starts a new batch of iterations locks a mutex and computes the chunk size for the current batch.
The subsequent threads simply read and reuse the already computed chunk size until the iterations in the batch have been scheduled. 
This requires mutex-based synchronization. 
\texttt{mFAC} avoids \AM{such costly} synchronization by involving more computation. 
Specifically, in \texttt{mFAC}, a shared counter is \emph{atomically} incremented so that threads identify the current batch number.
Hence, each thread calculates its \emph{own} next chunk size depending on the batch counter.
Depending on the synchronization and computation overheads in a given computing systems, one may use either \texttt{FAC} or \texttt{mFAC}.

\begin{figure}[!htb]
	\centering
	\includegraphics[width=0.94\columnwidth]{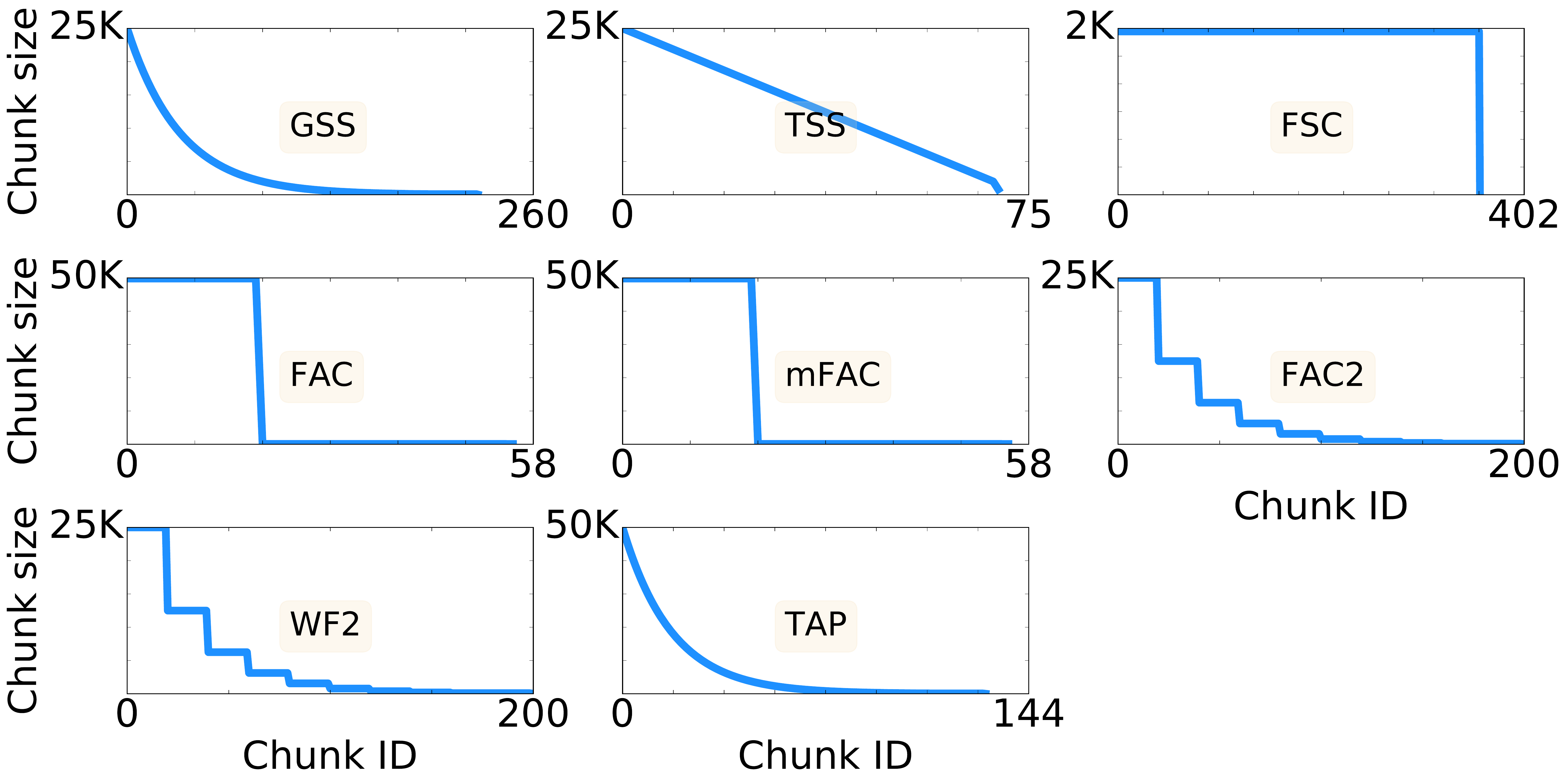}
	\caption{Progression of chunk sizes for \textit{dynamic} and \textbf{\textit{non-adaptive}} techniques for scheduling the main loop (\textbf{L1}) of SPHYNX with $1,000,000$ loop iterations on a 20-thread \xeon node and a chunk parameter of 97 loop iterations (Section~\ref{sec:factExperiments}).
		The chunk IDs are shown on the $x$ axis and the number of iterations per chunk on the $y$ axis. 
		}
	\label{fig:candynonadapt}
	\vspace{-0.1cm}
\end{figure}

\figurename~\ref{fig:candynonadapt} depicts an example of the chunk sizes calculated by the \textit{dynamic} and \textit{non-adaptive} scheduling techniques and their progression over the work requests for scheduling the iterations of the main loop (\textbf{L1}) of SPHYNX\footnote{\href{https://astro.physik.unibas.ch/en/people/ruben-cabezon/sphynx/}{astro.physik.unibas.ch/en/people/ruben-cabezon/sphynx/}} (more details in Section~\ref{sec:factExperiments}). 
\texttt{STATIC} and \texttt{SS} are not shown in \figurename~\ref{fig:candynonadapt} as their chunk size progression is constant (straight line), at the size of the chosen chunk parameter.
The chunk size progression for the \textit{dynamic} \textit{non-adaptive} scheduling techniques follows a decreasing chunk size pattern, wherein the next chunk of iterations is equal to or smaller than the previous.
One can note that not only the chunk sizes but also the total the number of chunks allocated varies among the scheduling techniques.
A small number of large chunk sizes may not mitigate load imbalance but incur a smaller scheduling overhead due to fewer scheduling operations while a greater number of smaller chunks may improve load balancing at the cost of increased scheduling overhead. 




\noindent \textbf{Dynamic and adaptive self-scheduling.}\\
Adaptive scheduling techniques regularly measure execution performance \textbf{during the application execution} and the scheduling decisions are taken based on this information. 
The adaptive scheduling techniques incur a higher scheduling overhead compared to non-adaptive techniques but are designed to outperform the non-adaptive ones in highly irregular execution environments.

\texttt{BOLD}~\cite{BOLD:1997} is a 'bolder' version of \texttt{FAC} and a further development of \texttt{TAP}. 
As with \texttt{TAP}, it uses the mean~$\mu$ and the standard deviation~$\sigma$ of the iteration execution times as well as an estimate of the scheduling overhead. 
The driving idea behind the BOLD strategy was to increase early chunk sizes such that scheduling overhead is reduced while considering the risk of potentially too large chunks of iterations. 

\texttt{AWF}~\cite{AWF:2003} is similar to \texttt{WF}  in that each thread executes variably-sized chunks of a given batch according to its relative processing weight.
The processing weight is updated during execution based on the performance of each thread.
\texttt{AWF} is devised for time-stepping applications and threads processing weights are updated at the end of each time-step.
Variants of \texttt{AWF}, namely \texttt{AWF-B} and \texttt{AWF-C}, relaxed this constraint by updating processing weights at the end of every batch and chunk execution, respectively.
Additional variants of \texttt{AWF}, namely \texttt{AWF-E} and \texttt{AWF-D}, are similar to \texttt{AWF-B} and \texttt{AWF-C}, respectively.
In addition, \texttt{AWF-E} and \texttt{AWF-D} take into account the overhead of scheduling in calculating the relative processing weights.


\texttt{AF}~\cite{AF:2000} is an adaptive DLS technique derived from \texttt{FAC}.
In contrast to \texttt{FAC}, \texttt{AF} \textit{learns} both $\mu$ and $\sigma$ for each computing resource during application execution to ensure \textit{full adaptability} to all factors that cause load imbalance.
\texttt{AF} adapts the chunk size during application execution based on the continuous updates of the mean loop iteration execution times $\mu$ and their standard deviation $\sigma$. 


\texttt{mAF} is our improved implementation of \texttt{AF}. 
In the original \texttt{AF} algorithm, the execution time of earlier loop iterations (from the same execution) are collected to calculate the next chunk size. 
The collected times only consider the execution time of the loop iterations themselves.
In \lbomp, \texttt{mAF} also considers the scheduling overhead.
Hence, \texttt{mAF} employs a more precise \textit{performance estimation} for calculating the chunk size, which is expected to lead to higher load balance and performance.

\begin{figure}[!htb]
	\centering
	\includegraphics[width=0.94\columnwidth]{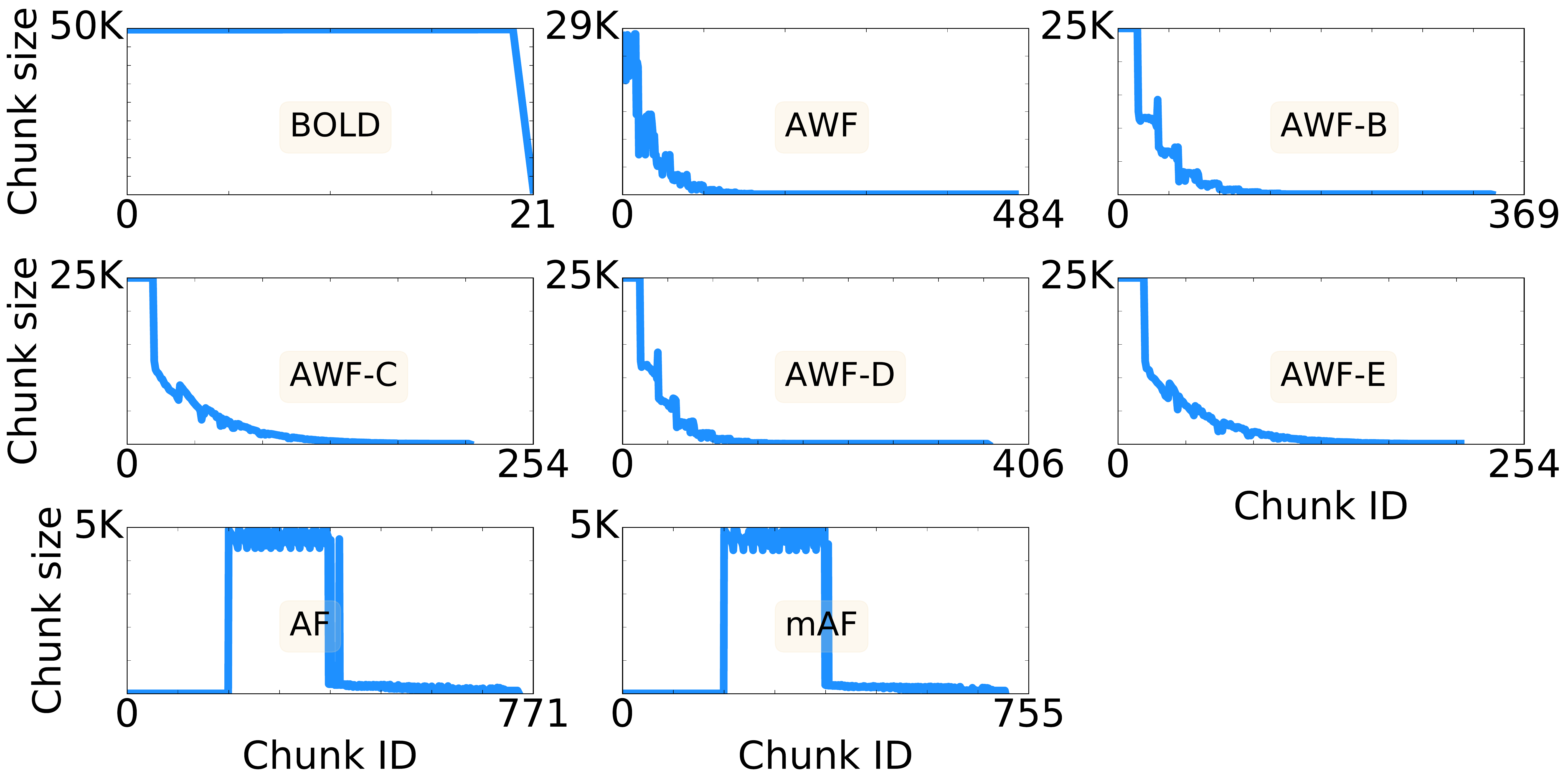}
	\caption{Progression of chunk sizes for \textit{dynamic} and \textbf{\textit{adaptive}} techniques for self-scheduling the main loop (\textbf{L1}) of SPHYNX with $1,000,000$ loop iterations on a 20-thread \xeon node and a \texttt{chunk} parameter of 97 loop iterations (Section~\ref{sec:factExperiments}).
	The chunk IDs are shown on the $x$ axis and number of loop iterations per chunk on the $y$ axis. 
		}
	\label{fig:candyadapt}
\end{figure}

\figurename~\ref{fig:candyadapt} shows the chunk sizes calculated by the \textit{dynamic} and \textit{adaptive} scheduling techniques and their progression over the work requests for scheduling the iterations of the main loop (\textbf{L1}) of \sphynx (more details in Section~\ref{sec:factExperiments}). 
The chunk sizes calculated by the adaptive scheduling techniques do not strictly decrease with each work request, but increase or decrease depending on the requesting thread's performance during execution.
\rwthree{\textbf{This is the quintessence of adaptive self-scheduling and load balancing: threads which require more time to compute receive less work while threads which compute faster receive more work.}}

\textit{Dynamic} and \textit{\textbf{adaptive}} self-scheduling and load balancing techniques will be critical for achieving performance on upcoming Exascale systems with heterogeneous architectures in which scheduling needs to dynamically adapt to threads executing on slower or faster processing \rwthree{units}. 
The observation from \figurename~\ref{fig:candynonadapt} also holds true for \figurename~\ref{fig:candyadapt} regarding the trade-off between fewer and larger chunks \textit{and} more and smaller chunks.




\subsection{Features for Performance Measurement}\label{sec:profile}
	

	\lbomp provides a number of features for performance measurements for target loops associated with the \openmp \texttt{schedule} clause.
	These features are crucial for the analysis of loop scheduling and load balancing. 
	
	\textbf{\textit{Thread execution time}.} This feature reports the threads execution times per loop execution instance. This information is important for the estimation of load imbalance in a parallel loop.
	This feature is enabled by defining the environment variable \texttt{KMP\_TIME\_LOOPS} and declaring the path where the measured performance data will be stored. 
	

	The \textit{thread execution time} feature enables the calculation of \mbox{well-known} load imbalance metrics such as \emph{coefficient of variation} (\cov)~\cite{FAC:1992} and \emph{percent imbalance} (\percentimbalance)~\cite{derose2007detecting}. 
	The \cov and \percentimbalance equations are defined in Table~\ref{table:exp}, where $T_{par}^{loop}$ denotes the parallel execution time of the loop. 
	
	
	

	In this work, these metrics are calculated based on the data measured with \lbomp for {\textit{individual} \openmp loops} and presented later in Section~\ref{sec:experiments}.

	\textbf{\textit{Chunk information}.} \lbomp collects and stores the calculated chunk sizes for each thread in each scheduling round. 
	This functionality can be enabled by setting the environment variable \texttt{KMP\_PRINT\_CHUNKS} to $1$.
	The collected information is stored at the location defined in \texttt{KMP\_TIME\_LOOPS} (see above).
    
    A detailed analysis of the chunk sizes calculated by each scheduling technique for given \openmp loops is fundamental for understanding their performance. We used this feature for the in-depth performance analysis of the impact of the chunk parameter on all scheduling techniques described in Sections~\ref{sec:minchunk} and~\ref{sec:calcchunk}.

	\textbf{\textit{Statistical information about loop iterations execution times}.} 
	
	\noindent\lbomp provides a \textit{profiling} feature that collects the mean of loop iterations execution times ($\mu$) and their standard deviation $\sigma$. 
	These measurements are required by the \texttt{FSC}, \texttt{FAC}, \texttt{TAP}, and \texttt{BOLD}  scheduling techniques.
	The \textit{profiling} feature relieves the (non-expert) user from the burden of collecting such profiling information.
	
	This feature can be enabled by defining \texttt{schedule(runtime)} in the target loop, exporting \texttt{OMP\_SCHEDULE=profiling}, and setting the environment variable \texttt{KMP\_PROFILE\_DATA} to the path where the profiling data will be stored. 

\subsection{Load Balancing Applications with \lbomp}\label{sec:howtouselb4omp}

\begin{figure}[!htb]
	\vspace{-0.2cm}
	\centering
	\includegraphics[width=0.85\linewidth]{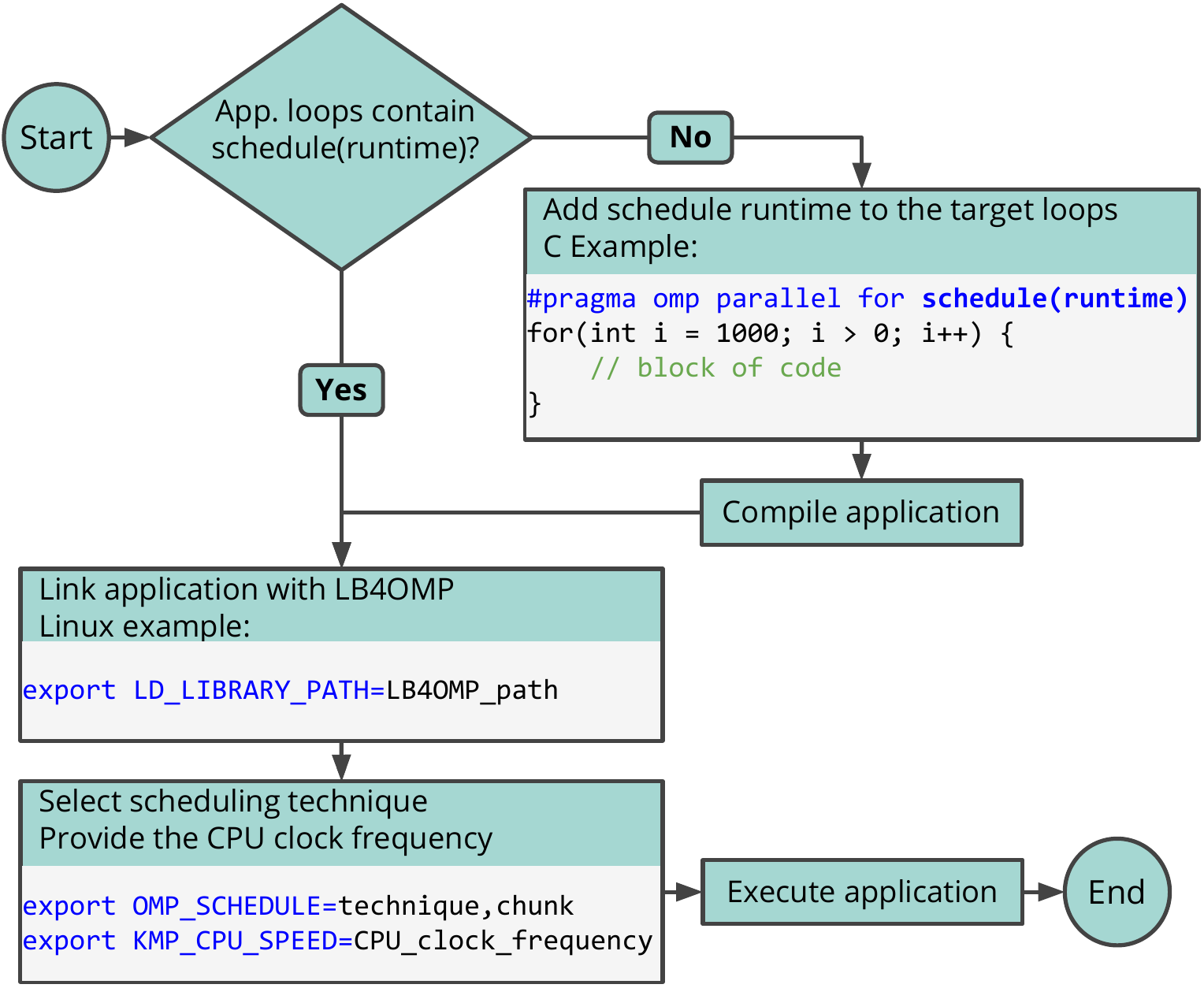}
	\caption{Workflow for dynamic load balancing the execution of \openmp applications using the scheduling techniques in \lbomp. 
	}
	\label{fig:lb4ompusage}
\end{figure}

The use of \lbomp with an \openmp application is straightforward and illustrated in \figurename~\ref{fig:lb4ompusage}.
First, one must ensure that the target \openmp loops in the application contain the \texttt{schedule(runtime)} clause.
If that is the case, no other changes are required and there is no need to recompile the code.
Otherwise, one needs to change (or add, if the loop structure permits) the existing scheduling clause to \texttt{runtime} in all target loops and recompile the application.
Next, one needs to add the path to the compiled \lbomp to the environment variable that the linker uses to load dynamic and shared libraries (e.g., \texttt{LD\_LIBRARY\_PATH} on Linux/Unix systems).
%
The workflow in~\figurename~\ref{fig:lb4ompusage} is almost independent of the target system. 
The only system-related parameter required by \lbomp is the host CPU clock frequency. 
This is passed to \lbomp via the environment variable \texttt{KMP\_CPU\_SPEED} as an integer variable in MHz.

\rwthree{
The adaptive scheduling techniques in \lbomp use low overhead cycle counters as \texttt{RDTSCP} to measure the execution time of previous chunks of iterations.
We use the clock frequency to convert the cycles into time (which are the values expected by the formulas of those techniques).
One may argue that the static value defined by the user in \texttt{KMP\_CPU\_SPEED} may be inaccurate for a number of modern processors that allow dynamic clock frequency change during execution.
The measured performance of the threads is relative to each other, so variations in clock frequency during execution do not affect the chunk calculation.
Work is ongoing to automate the process of collecting the clock speed for use in \lbomp.
}

Applications with multiple loops may need to employ different scheduling techniques in different parts of the code.  
\lbomp uses the \texttt{schedule(runtime)} option available in \openmp and, therefore, the scheduling technique selected by the user is read from the environment variable \texttt{OMP\_SCHEDULE}. 
To select different scheduling techniques for an application in different parts of the code, one can use the function specified in the \openmp standard \texttt{omp\_set\_schedule(omp\_sched\_t kind, int chunk\_size)}\footnote{\href{www.openmp.org/spec-html/5.0/openmpsu121.html}{www.openmp.org/spec-html/5.0/openmpsu121.html}}.
This is used to update the scheduling technique specified with \texttt{OMP\_SCHEDULE} during execution.
One can also export environment variables directly from the application code to update the configuration of \lbomp and, if preferred, to update the scheduling technique itself.

\section{Performance Results and Discussion}\label{sec:experiments}

We use \rwthree{three applications, two microbenchmarks}, and three computing node types to evaluate the performance of the existing in LLVM \openmp RTL and newly implemented DLS techniques in \lbomp (see Table~\ref{table:exp}).

\rwthree{
	The applications 352.nab\footnote{352.nab: \href{https://www.spec.org/omp2012/Docs/352.nab.html}{www.spec.org/omp2012/Docs/352.nab.html}}, \sphynx, and the microbenchmark \dist\footnote{\dist: \href{https://drive.switch.ch/index.php/s/XIhieSdmoRuLcRR}{drive.switch.ch/index.php/s/XIhieSdmoRuLcRR}} were selected since they contain imbalanced and computationally-intensive loops,
	which are the most promising optimization targets for improved performance with dynamic (and adaptive) self-scheduling techniques~\cite{ciorba:2018}.
	The application \gromacs~\cite{gromacsPaper} and the microbenchmark \stream\footnote{\stream: \href{http://www.cs.virginia.edu/stream/ref.html}{www.cs.virginia.edu/stream/ref.html}} were selected to provide an overview of the scheduling overhead, ccNUMA effects, and locality loss incurred by the scheduling techniques in \lbomp.
	These loops are balanced, of low arithmetic intensity, and mainly perform memory operations which stresses the possibly negative effects of dynamic (and adaptive) self-scheduling.
}

We will use the following notation to specify details regarding the applications and systems. 
$T$ denotes number of \mbox{time-steps}, 
$L$ the IDs of loop with modified \texttt{schedule} clauses, and 
\timenoloop the time spent by the application outside of loops. 
The loops for which we modify the \texttt{schedule} clause were parallel and not nested\footnote{\lbomp can schedule nested and non-nested parallel loops with independent iterations.}. 

\subsection{Design of Factorial Experiments}\label{sec:factExperiments}

	Table~\ref{table:exp} presents the design of the factorial experiments needed for the extensive performance analysis campaign.
	
	For 352.nab, part of the SPEC OMP 2012 benchmark suite, we used the \textit{reference} input size. 
    For \sphynx, the Evrard Collapse test case was performed with $1,000,000$ particles.
	For \gromacs, the input size used in the experiments was the \textit{Test Case B} taken from the Unified European Application Benchmark Suite (UEABS)\footnote{UEABS: \href{https://repository.prace-ri.eu/git/UEABS/ueabs}{repository.prace-ri.eu/git/UEABS/ueabs}}.
	\rwthree{
	The \stream microbenchmark was executed with its default array size of $80,000,000$ elements (doubles), memory per array $610.4$ MB, which required a total of $1831.1$ MB in memory.
	\dist is synthetic microbenchmark used to show how the scheduling techniques react to different statistical loop workload distributions across iterations. 
	Each loop of \dist follows a different workload distribution as indicated in Table~\ref{table:exp}.
	}
	
	Each experiment was repeated 5 times \rwthree{(\stream was repeated 20 times)} and the average execution time \rwthree{or memory bandwidth (in MB/s) for \stream} is reported. 
	The applications 352.nab, \sphynx, and \gromacs are \mbox{time-stepping} simulations. 
	The \textit{computationally-intensive loops} with modified \texttt{schedule} clause from each application and microbenchmark are indicated in Table~\ref{table:exp}.
	The applications and the \lbomp library were compiled with the Intel compiler version 19.0.1.144. 
	The characteristics of the computing systems are also indicated in Table~\ref{table:exp}.
	
	Throughout the performance analysis campaign, all scheduling techniques used the \textit{default chunk parameter} (1~loop iteration).
	We used the \textit{thread execution time} \lbomp feature (Section~\ref{sec:profile}) and measured the loop execution and threads finishing times per loop to derive load imbalance. 
	
	\begin{table*}[!htb]
	\caption{Criteria used for the design of factorial experiments, resulting in a total of $4,826$ experiments.}
	\begin{center}
		\vspace{-0.2cm}\resizebox{0.96\textwidth}{!}{
			\begin{tabular}{l|l|l|l}
				\hline
				\multicolumn{2}{l|}{\textbf{Factors}}                                                      & \textbf{Values}                                         & \textbf{Properties}                                    \\ \hline
				\multicolumn{2}{l|}{}             & SPEC OMP 2012 352.nab             & \begin{tabular}[c]{@{}l@{}}$N$ = $44,794$ $|$ $T$ = $1,002$ $|$ Total loops  = $13$ $|$ Modified loops = $7$\\ 
				\end{tabular}\\ \cline{3-4} 
			
				\multicolumn{2}{l|}{Applications}     & SPHYNX Evrard Collapse       & \begin{tabular}[c]{@{}l@{}}$N$ = $1,000,000$ $|$ $T$ = $20$ $|$ Total loops  = $37$ $|$ Modified loops = $2$\\  
		         \end{tabular}    \\ \cline{3-4} 
	            
	            \multicolumn{2}{l|}{}             & GROMACS             &    \begin{tabular}[c]{@{}l@{}}$N$ = $3,316,463$ $|$ $T$ = $10,000$ $|$ Total loops  = $90$ $|$ Modified loops = $1$\\ 
			    	\end{tabular}    \\ \hline
	            
			\multicolumn{2}{l|}{\rwthree{Microbenchmarks}}       & \rwthree{\stream}         & \begin{tabular}[c]{@{}l@{}}\rwthree{$N$ = $80,000,000$ $|$ $T$ = $1$ $|$ Total loops  = $4$ $|$ Modified loops = $4$}\\ 
				\rwthree{$L0$ (copy): $a(i) = b(i)$, bytes/iteration = 16, FLOP/iteration = 0} \\
				\rwthree{$L1$ (scale): $a(i) = q*b(i)$, bytes/iteration = 16, FLOP/iteration = 1} \\
				\rwthree{$L2$ (add): $a(i) = b(i) + c(i)$, bytes/iteration = 24, FLOP/iteration = 1}\\
				\rwthree{$L3$ (triad): $a(i) = b(i) + q*c(i)$, bytes/iteration = 24, FLOP/iteration = 2}
				\end{tabular}     \\  \cline{3-4} 
			    
			    \multicolumn{2}{l|}{}       & \rwthree{\dist}        & \begin{tabular}[c]{@{}l@{}}\rwthree{$N$ = $1,000$ $|$ $T$ = $1$ $|$ Total loops  = $5$ $|$ Modified loops = $5$}\\ 
				\rwthree{$L0$ (constant): $2.3\times10^8$ FLOP per iteration, }\\ 
                \rwthree{$L1$ (uniform): [$10^3, 7\times10^8$] FLOP per iteration, }\\
                \rwthree{$L2$ (normal): $\mu = 9.5\times10^8$ FLOP, $\sigma = 7\times10^7$ FLOP, $[6\times10^8, 1.3\times10^9]$ FLOP per iteration, }\\
                \rwthree{$L3$ (exponential): $\lambda = 1/3\times10^8$ FLOP, $[948, 4.5\times10^9]$ FLOP per iteration, }\\
                \rwthree{$L4$ (gamma): $k = 2$, $\theta= 10^8$ FLOP, $[4.1\times10^6, 2.7\times10^9]$ FLOP per iteration,}
				\end{tabular}     \\  \hline 


				&                                          & \texttt{static} (\texttt{STATIC})                                                                                                                                                                                                                                                                  & Straightforward parallelization                                                                                                                                                                               \\ \cline{3-4} 
				& \multirow{-2}{*}{OpenMP standard}        & \texttt{guided} (\texttt{GSS}), \texttt{dynamic,1} (\texttt{SS}) &                                                                                                                                                                                                               \\ \cline{2-3}
				& \multicolumn{1}{r|}{OpenMP non-standard} & \texttt{TSS} &                                                                                                                                                                                                               \\ \cline{2-3}
				&                                          & \texttt{FSC}, \texttt{FAC}, \texttt{FAC2}, \texttt{TAP}, \texttt{WF2}, \texttt{mFAC} & \multirow{-3}{*}{\textbf{Dynamic} and \textit{\textbf{non-\textit{adaptive}}} self-scheduling techniques} \\ \cline{3-4} 
				\multirow{-5}{*}{\begin{tabular}[c]{@{}l@{}}Scheduling\\ techniques\end{tabular}} & \multirow{-2}{*}{\textbf{LB4OMP}} & \texttt{BOLD}, \texttt{AWF}, \texttt{AWF-B}, \texttt{AWF-C}, \texttt{AWF-D}, \texttt{AWF-E},  \texttt{AF}, \texttt{mAF} & \textbf{Dynamic} and \textit{\textit{\textbf{adaptive}}} self-scheduling techniques                                                                                                                                                               \\ \hline
				\multicolumn{2}{l|}{Chunk parameter} & \begin{tabular}[c]{@{}l@{}}$N/(2P)$, $N/(4P)$, $N/(8P)$, $N/(16P)$, ..., 1\\ Fewest number of chunk parameter tested: \\ 7 for 352.nab on node \knl with hyperthreading\\ Largest number of chunk parameter tested: \\ 16 for SPHYNX on node \daint without hyperthreading\end{tabular} & \begin{tabular}[c]{@{}l@{}} For \texttt{dynamic} and \texttt{static} the chunk parameter denotes the fixed amount\\ of iterations in every chunk\\ \\For all others DLS techniques, the chunk parameter represents the smallest \\chunk size a thread can obtain with a given self-scheduling technique\end{tabular} \\ \hline
				\multicolumn{2}{l|}{} & \xeon & \begin{tabular}[c]{@{}l@{}}Intel Broadwell E5-2640 v4 (2 sockets, 10 cores each)\\ $P$ =  20 without hyperthreading, $P$ = 40 with hyperthreading\\ Pinning: \texttt{OMP\_PLACES}=cores, \texttt{OMP\_PROC\_BIND}=close\end{tabular} \\ \cline{3-4} 
				\multicolumn{2}{l|}{Computing nodes} & \knl & \begin{tabular}[c]{@{}l@{}}Intel Xeon Phi KNL 7210 (1 socket, 64 cores)\\ $P$ =  64 without hyperthreading, $P$ = 256 with hyperthreading\\ Pinning: \texttt{OMP\_PLACES}=cores, \texttt{OMP\_PROC\_BIND}=close\end{tabular}         \\ \cline{3-4} 
				\multicolumn{2}{l|}{\multirow{-4}{*}{}} & \daint & \begin{tabular}[c]{@{}l@{}}Intel Xeon E5-2690 v3 (1 socket, 12 cores)\\ $P$ = 12 without hyperthreading, $P$ = 24 with hyperthreading\\ Pinning: \texttt{OMP\_PLACES}=cores, \texttt{OMP\_PROC\_BIND}=close\end{tabular}             \\ \hline
				 
				  \multicolumn{4}{l}{}                \\ \hline
			 	
				      \multicolumn{2}{l|}{}                                         & Performance per loop                 &    Parallel loop execution time  $T_{par}^{loop}$ \\ \cline{3-4}
				  \multicolumn{2}{l|}{{Metrics}}            & Load imbalance per loop            &      \begin{tabular}[c]{@{}l@{}} \cov $= \sigma/\mu$ \\ \percentimbalance $ = \frac{T_{par}^{loop}-\mu}{T_{par}^{loop}}\times \frac{P}{P-1} \times 100\%$ \end{tabular} \\ \hline
			\end{tabular}
		}\vspace{-0.2cm}
		\label{table:exp}
	\end{center}
\end{table*}
\begin{figure*}[!htb]
		\centering
		\vspace{-1.8cm}
		\adjustbox{width=1.2\textwidth}{
\hspace{-2cm}		\begin{minipage}{1.3\textwidth}
		
		\subfigure{
			\includegraphics[width=.3\textwidth] {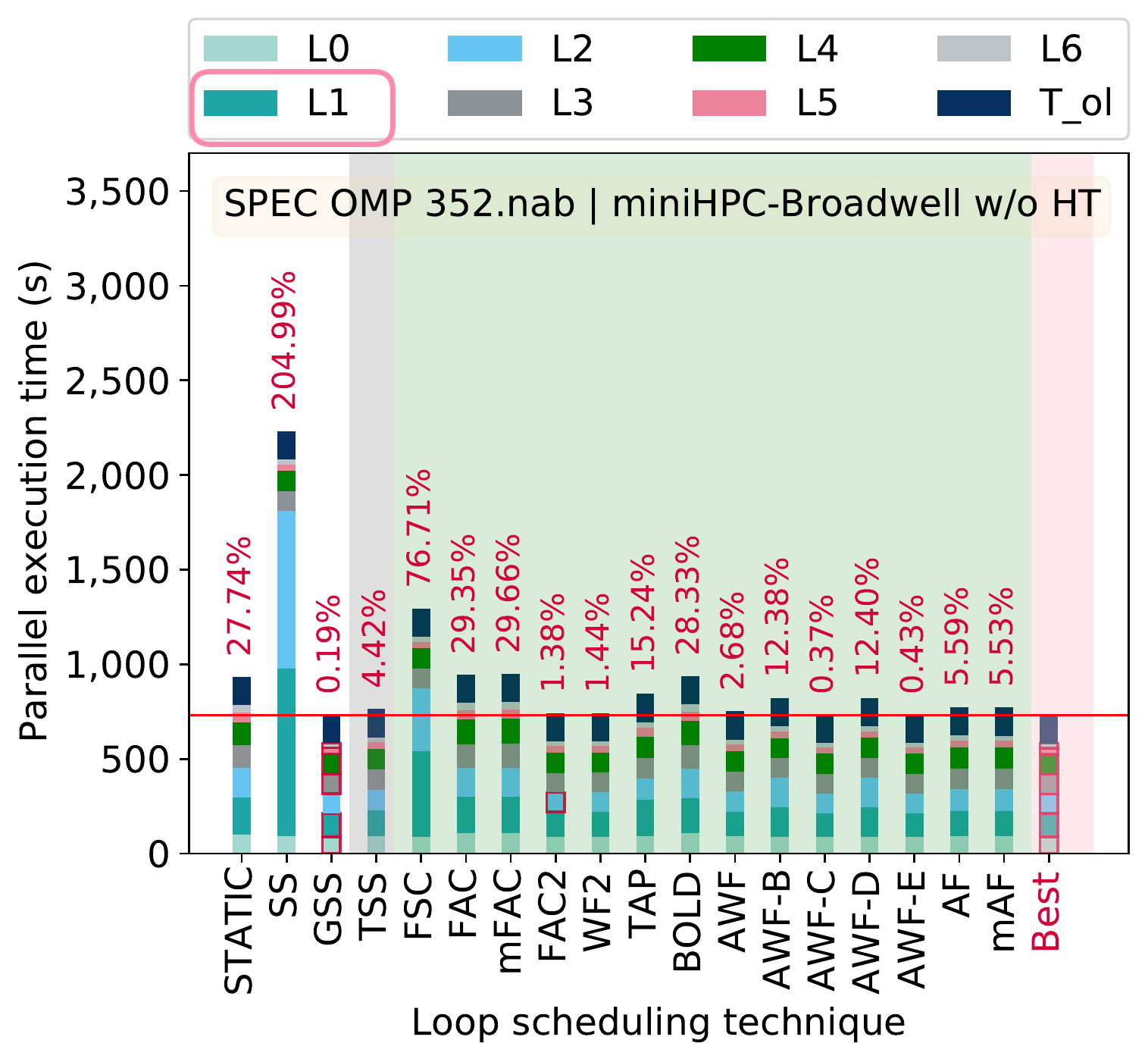}
			\label{fig:nabXeon}
		}
		\subfigure{
			\tcbox[colframe=highlight,
			colback=white, top=0pt,left=0pt,right=0pt,bottom=0pt]{\includegraphics[width=.3\textwidth]{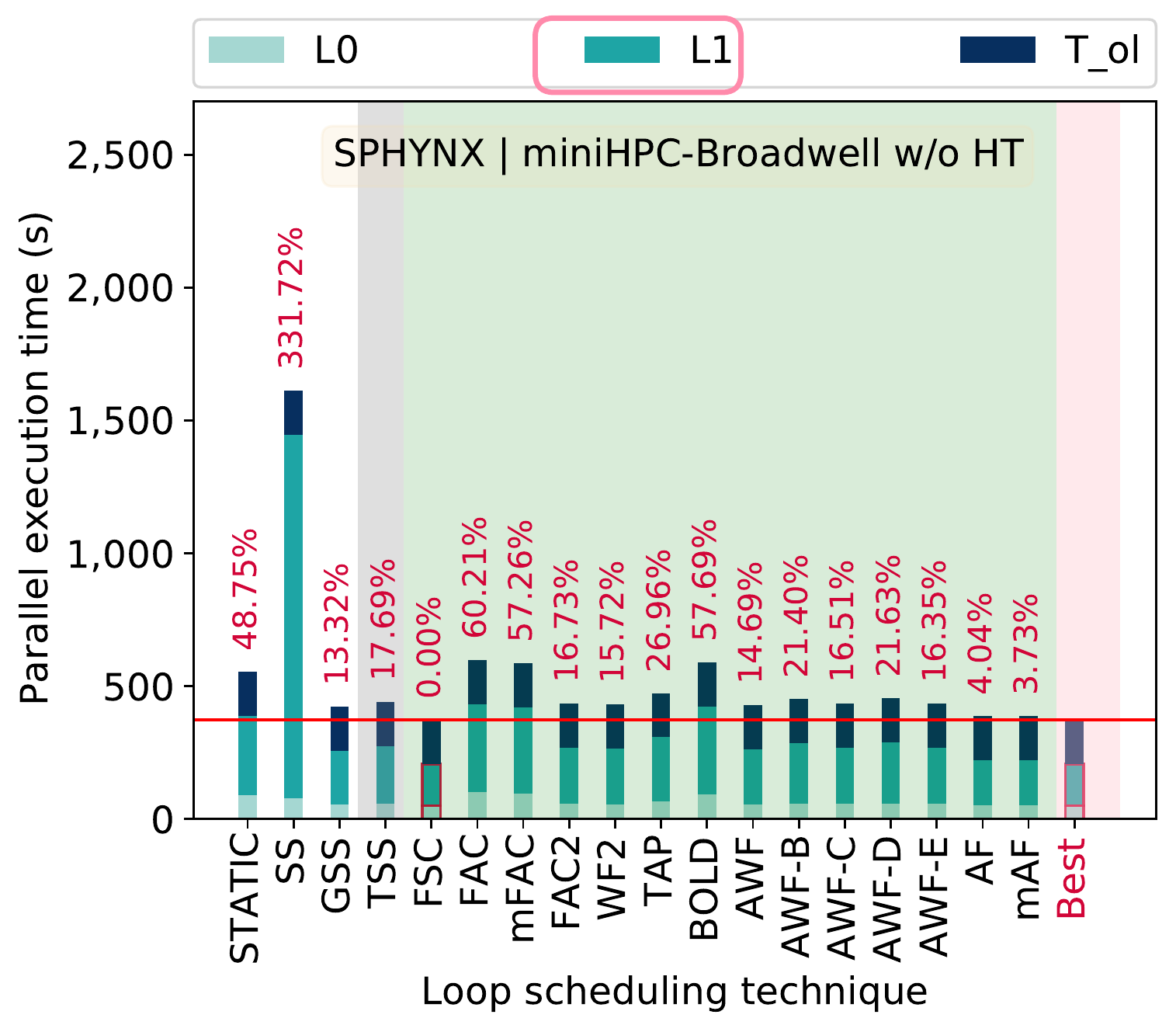}}
			\label{fig:sphynxXeon}
		}
		\subfigure{
			\includegraphics[width=.3\textwidth]{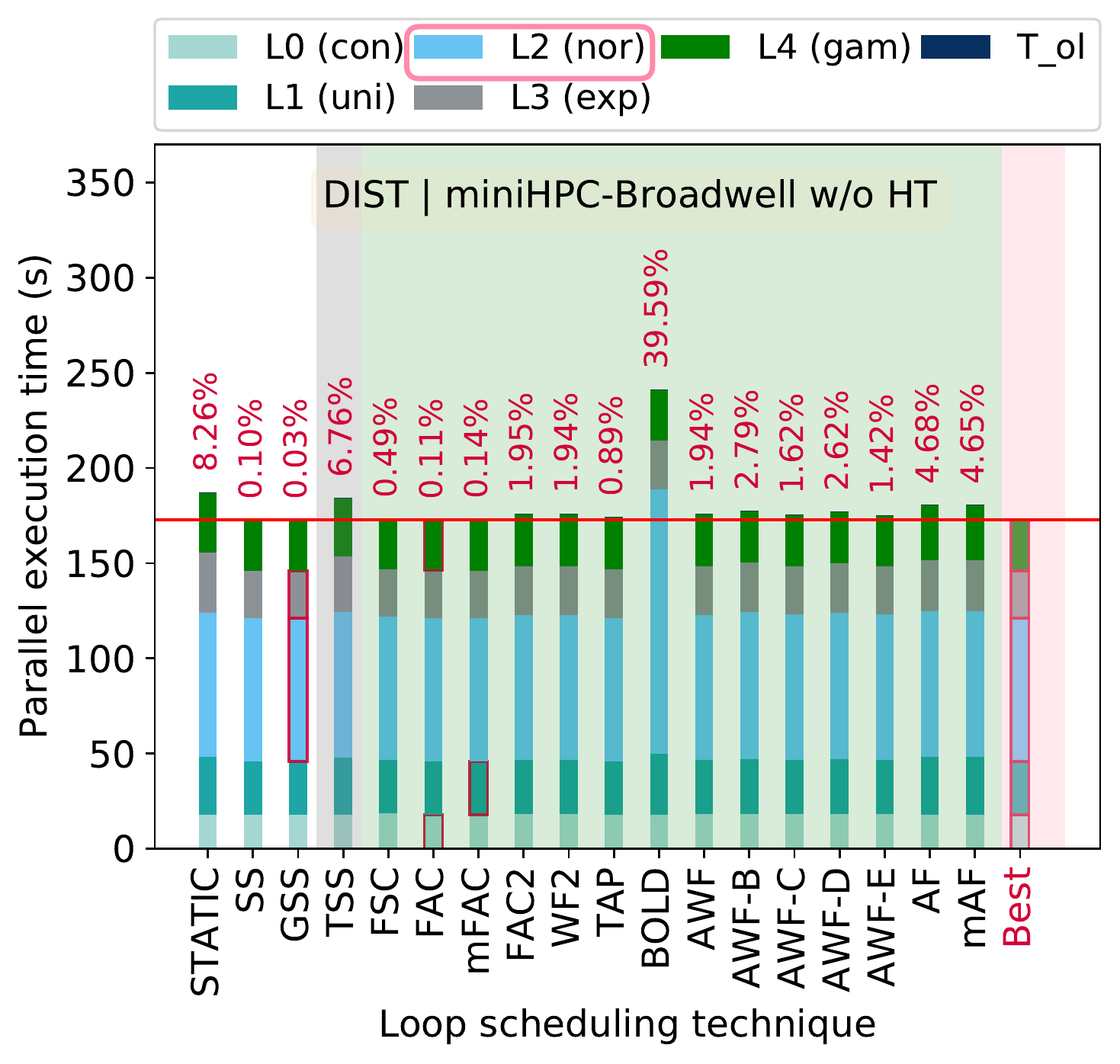}
			\label{fig:sphexaXeon}
		}\vspace{-0.3cm}\\
		\subfigure{
			\includegraphics[trim=0cm 0cm 0cm 2.05cm,  clip=true, width=.3\linewidth]{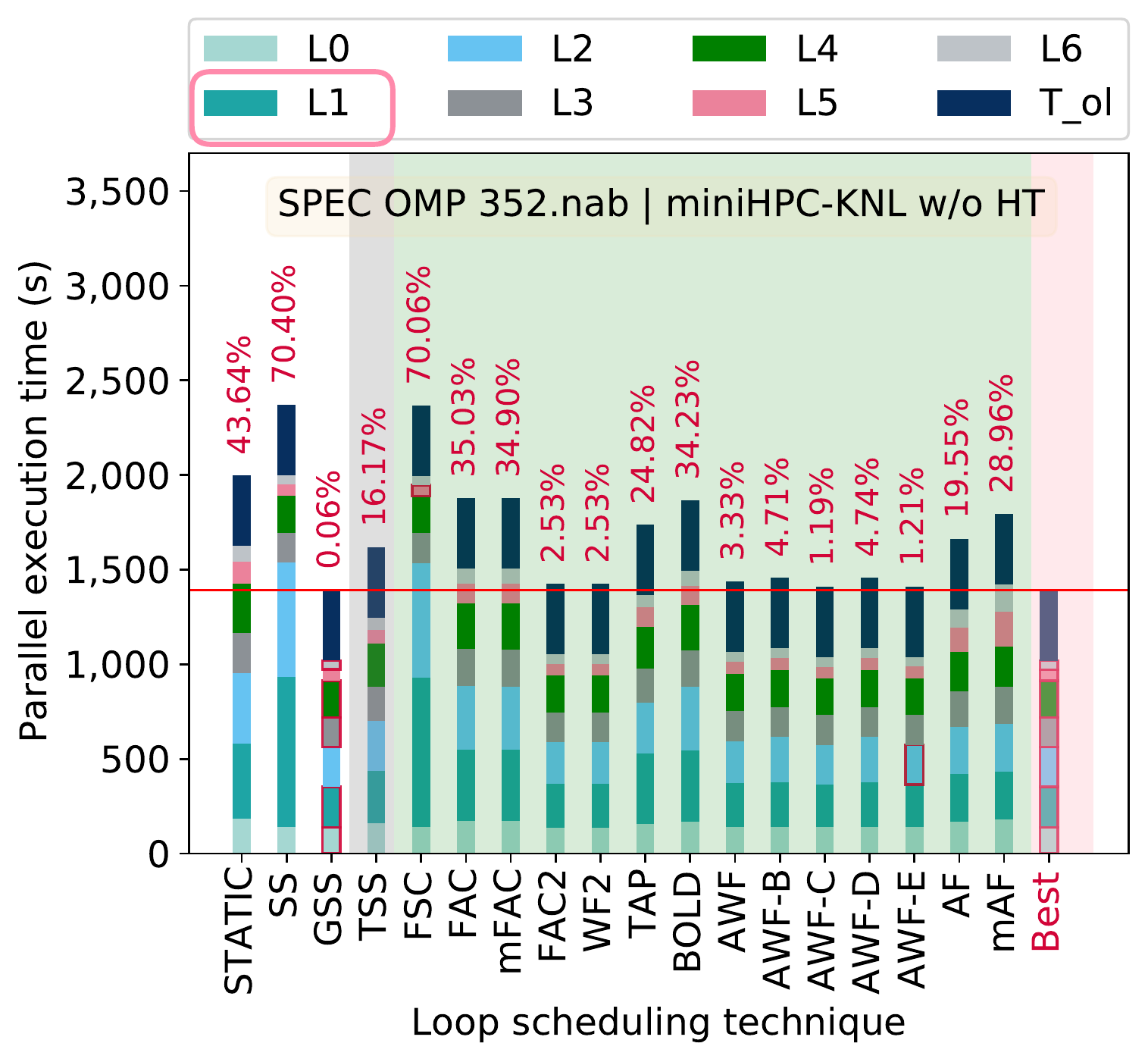}
			\label{fig:nabKNL}
		}
		\subfigure{
			\tcbox[colframe=highlight,
			colback=white, top=0pt,left=0pt,right=0pt,bottom=0pt]{\includegraphics[trim=0cm 0cm 0cm 1.3cm,  clip=true, width=.3\linewidth]{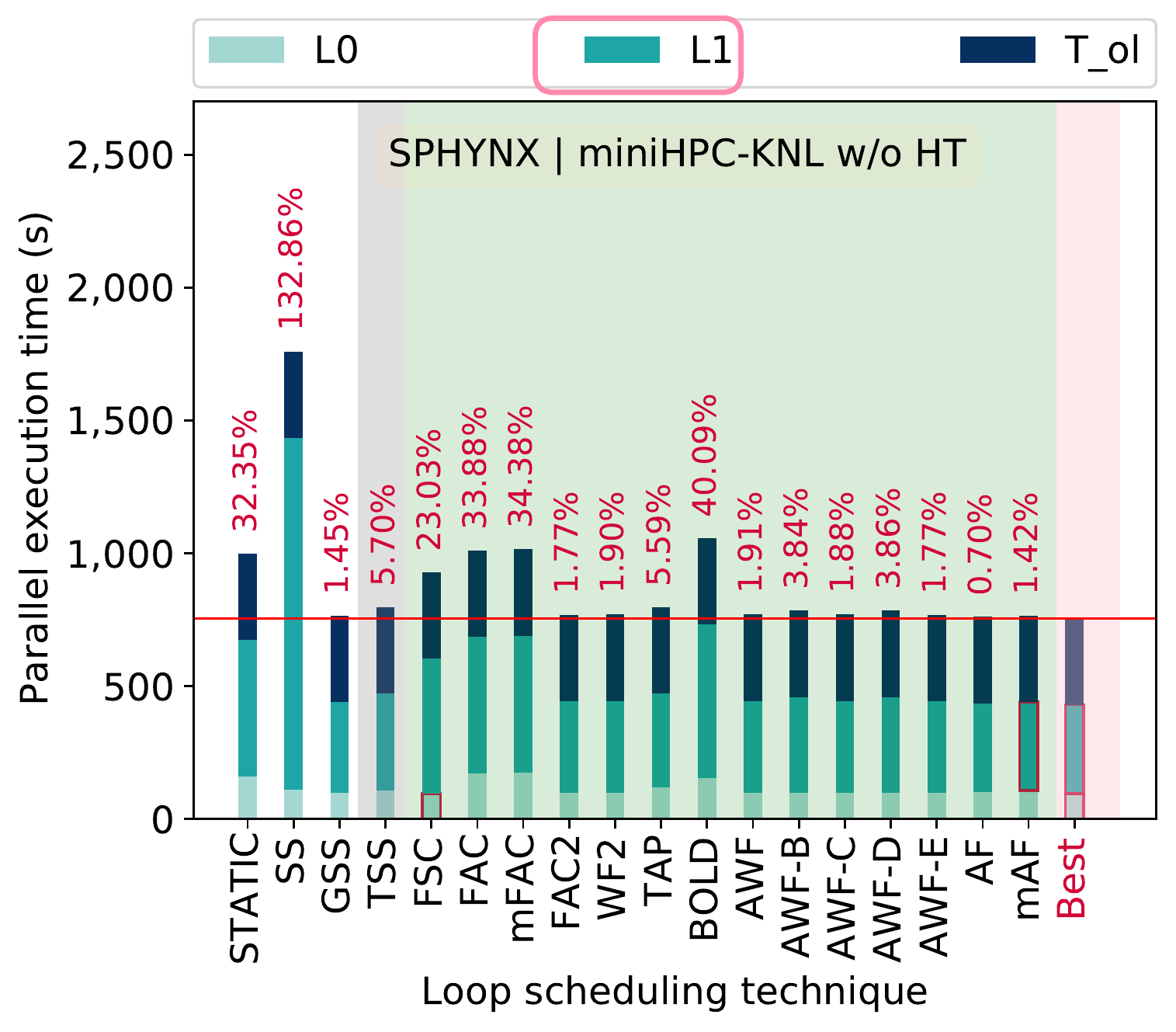}}
			\label{fig:sphynxKNL}
		}
		\subfigure{
			\includegraphics[trim=0cm 0cm 0cm 1.8cm,  clip=true, width=.3\linewidth]{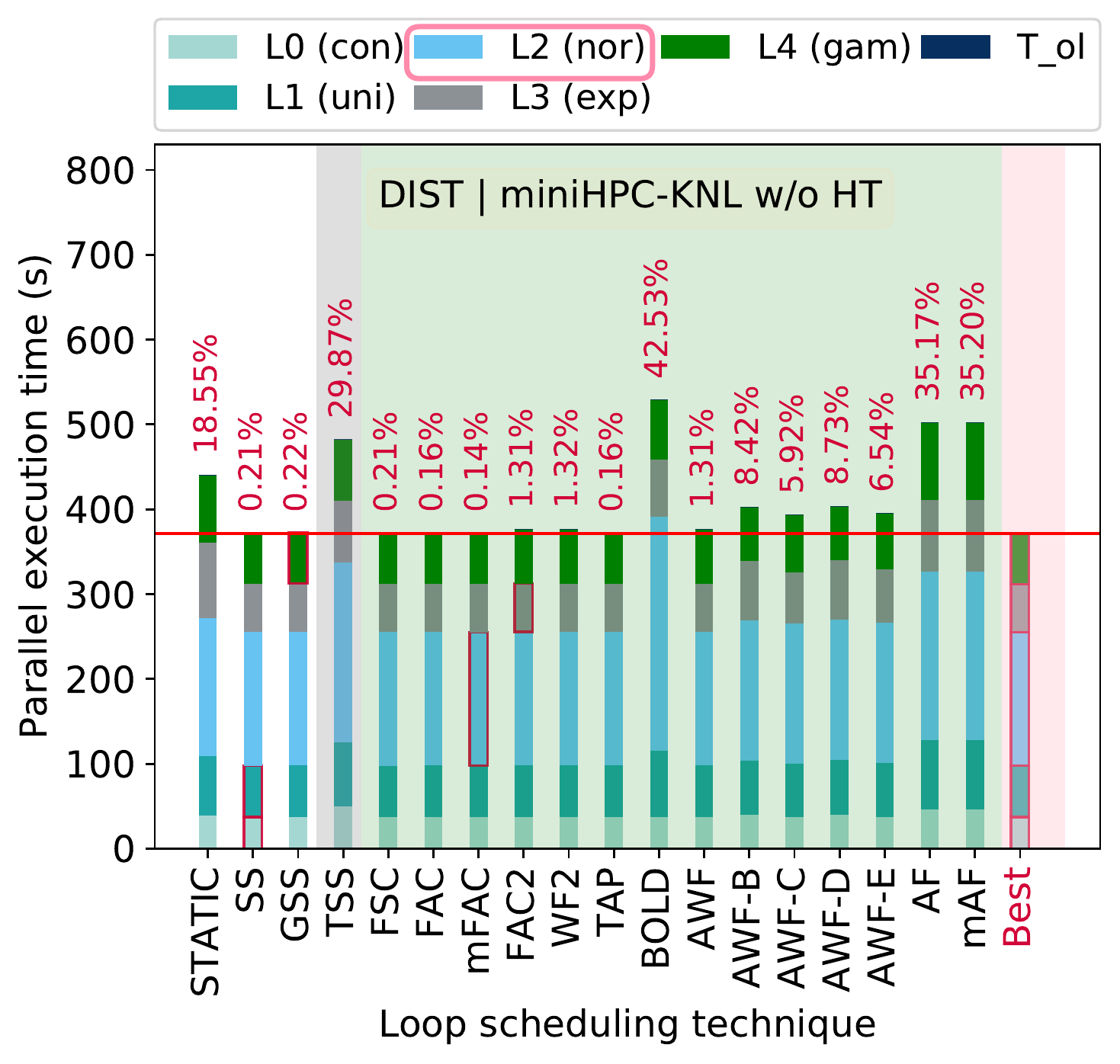}
			\label{fig:sphexaKNL}
		}\vspace{-0.3cm}\\
		\subfigure{
			\includegraphics[trim=0cm 0cm 0cm 2.05cm,  clip=true, width=.3\linewidth]{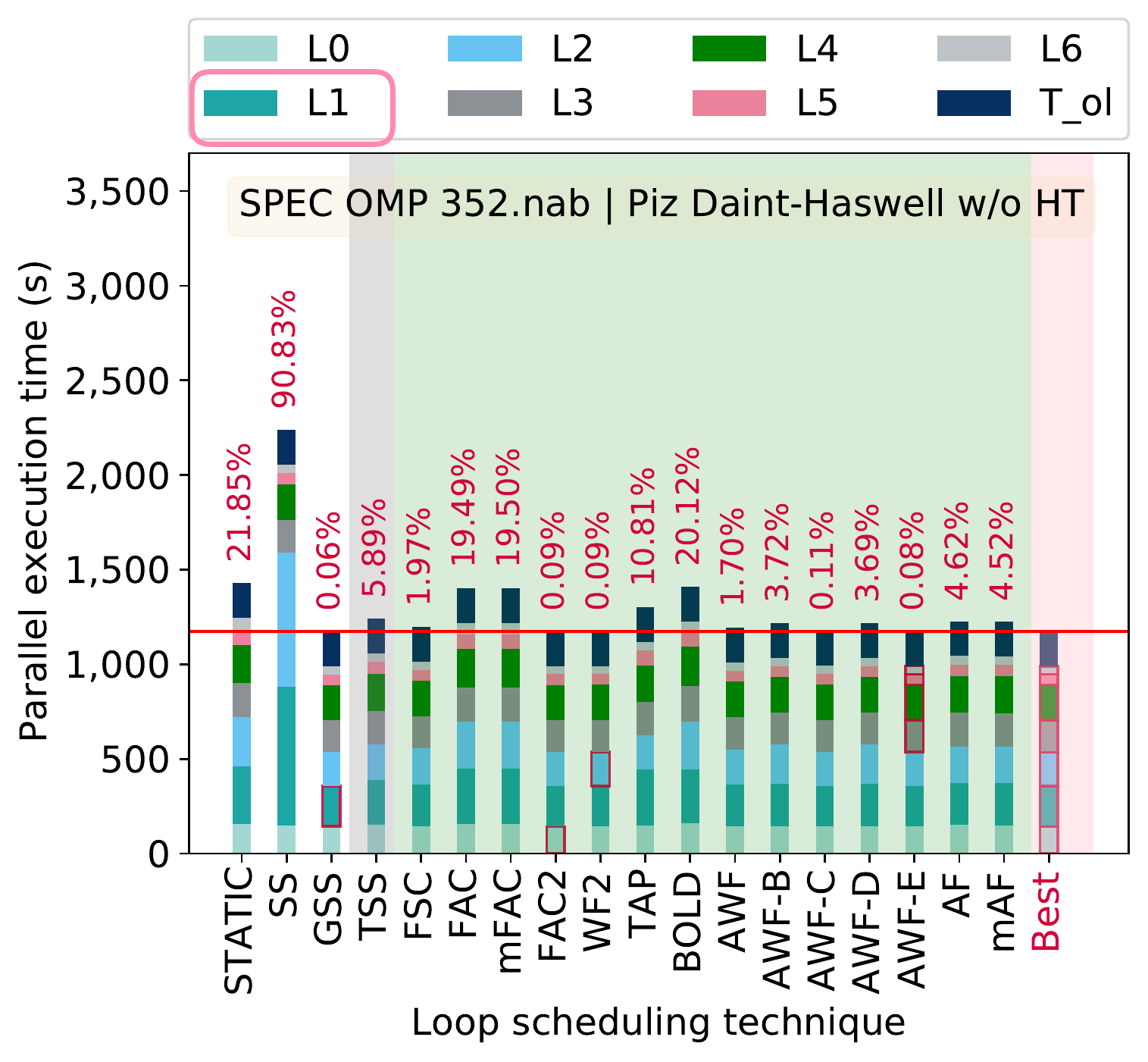}
			\label{fig:nabdaint}
		}
		\subfigure{
			\tcbox[colframe=highlight,
			colback=white, top=0pt,left=0pt,right=0pt,bottom=0pt]{\includegraphics[trim=0cm 0cm 0.2cm 1.3cm,  clip=true, width=.3\linewidth]{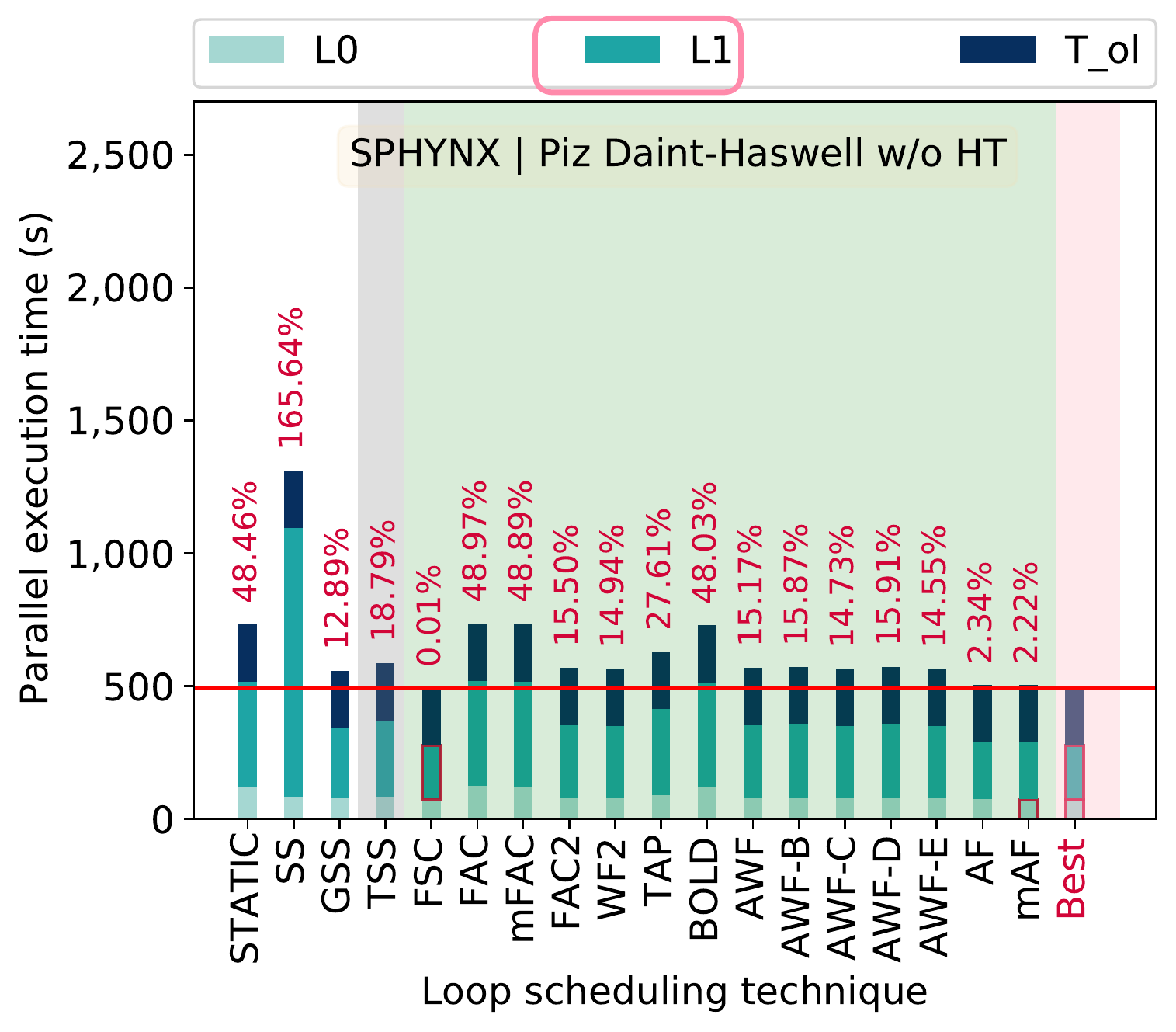}}
			\label{fig:sphynxdaint}
		}
		\subfigure{
			\includegraphics[trim=0cm 0cm 0cm 1.8cm,  clip=true, width=.3\linewidth]{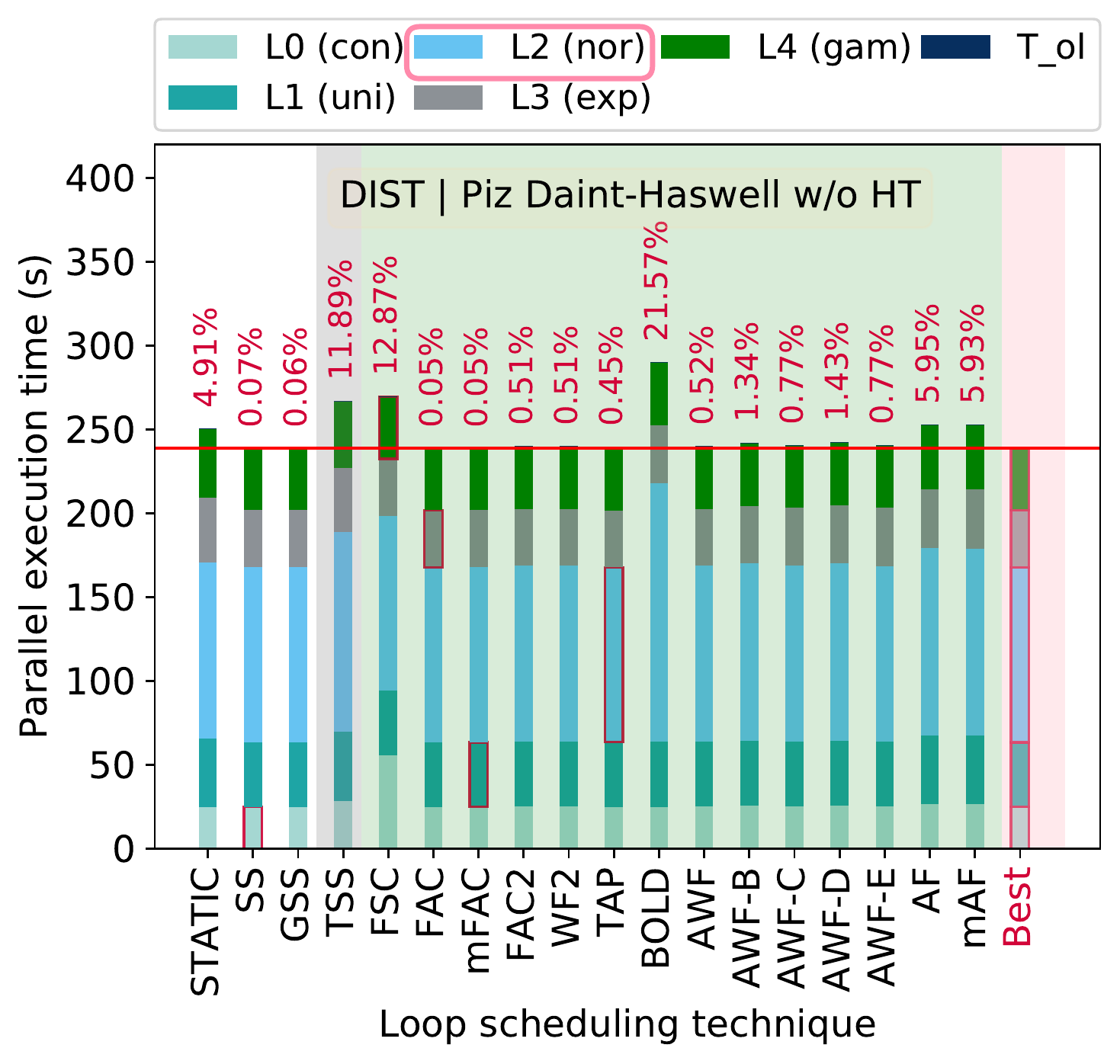}
			\label{fig:sphexadaint}
		}\\
	  	
		\vspace{-0.3cm}
		\caption[]{
		Average parallel execution time for each modified loop in 352.nab, \sphynx, and \dist with the default chunk parameter executing on all node types without hyperthreading.
		The $x$ axis shows the DLS techniques, while the $y$ axis presents the parallel execution time of each modified loop ($L$).
		The most time consuming loop of each application is highlighted in the legend.
        The red rectangles encompassing the bars represent the \emph{best} performing scheduling technique for a given loop.
        On the $x$ axis, the \texttt{Best} presents the shortest achievable execution time by selecting the combination of all individually highest performing techniques per loop together. 
        The background color highlights: in white, the \openmp standard DLS techniques, in gray, the non-standard technique that was already implemented in the LLVM \openmp RTL, in green \lbomp, and in dark pink the \texttt{Best} combination of techniques.
		The percentages denote performance degradation due to executing the applications with a single DLS technique vs. using the \texttt{Best} combination. 
		The plots highlighted in light pink will be further explored in the next sections.
		}
	\end{minipage}
}
		\label{fig:allresults}
	\end{figure*}

\subsection{Performance Analysis}\label{sec:perfeval}

	The goal of this performance analysis campaign is to examine which DLS technique provides the \emph{highest} performance and \emph{lowest} load imbalance for each application's loop scheduled with \lbomp.


%



\figurename~\ref{fig:allresults} shows the average parallel execution time for each modified loop in 352.nab, \sphynx, and \dist with the default chunk parameter executing on all node types without hyperthreading.
In \figurename~\ref{fig:allresults}, the \texttt{Best} combination of scheduling techniques varies greatly between applications and systems, and outperforms every single technique in most cases. 
This \emph{reinforces} the need for additional scheduling options in \openmp~\cite{ciorba:2018} since the \texttt{Best} combination commonly includes the techniques implemented in \lbomp. 

From the results of the \textit{dynamic} and \textit{non-adaptive} scheduling techniques in \figurename~\ref{fig:allresults}, we observe that \texttt{FAC2} and \texttt{GSS} presented fairly high performance in almost all experiments, despite the performance for \sphynx on node \xeon and \daint. 
\rwthree{Despite the high performance achieved by \texttt{TSS}, \texttt{FAC}, \texttt{mFAC}, and \texttt{TAP} for \dist, these techniques achieved low performance for the majority of other applications and systems.}
With profiling information, \texttt{FSC} calculated a proper chunk size achieving high performance in almost all experiments, despite the performance for \sphynx on node \knl, 352.nab on nodes \xeon and \knl, and \rwthree{\dist on \daint.}
The highest achieved performance improvement with a \textit{dynamic} and \textit{non-adaptive} scheduling technique was on \sphynx with \texttt{FSC} on \xeon outperforming \texttt{GSS}, the best standard technique in this case, by $13.32\%$.

The \textit{dynamic} and \textit{adaptive} loop scheduling techniques naturally add overhead. 
However, they 
adapt to application and system variations and heterogeneity without requiring profiling information. 
In \figurename~\ref{fig:allresults}, we observe that, except for \texttt{BOLD} (and for \texttt{AF}, \texttt{mAF} for 352.nab and \dist on \knl), all adaptive scheduling techniques consistently achieved high performance.
In numerous cases, the adaptive scheduling techniques are included in the \texttt{Best} combination. 
For example, \texttt{AF} and \texttt{mAF}, consistently presented high performance for all results with \sphynx, in which \texttt{mAF} is included in the \texttt{Best} combination for both \knl and \daint nodes.
The highest achieved performance improvement with a \textit{dynamic} and \textit{adaptive} scheduling technique was on \sphynx with \texttt{mAF} on \daint outperforming \texttt{GSS}, the best standard technique in this case, by $10.67\%$.


The results for \sphynx on node \xeon show that \texttt{AF} and \texttt{mAF} reasonably outperformed \texttt{GSS} by approximately $9.28\%$ and $9.59\%$ respectively. 
\sphynx executing on \xeon node also shows that \texttt{FSC} calculated a proper chunk size for both loops, obtaining the highest overall performance, outperforming \texttt{GSS} by $13.32\%$ and \texttt{mAF} by $3.73\%$. 
This behavior is consistent among the results on nodes of \knl and \daint (\figurename~\ref{fig:allresults}). 
In the following Section~\ref{sec:minchunk}, we further investigate the performance of the scheduling techniques for the most time-consuming loop of \sphynx, $L1$, while varying the chunk parameter.

\begin{figure}[!htb]
		\centering
	  		\subfigure{
	  			\includegraphics[trim=0.2cm 0.1cm 0.3cm 0cm,  clip=true, width=\linewidth]{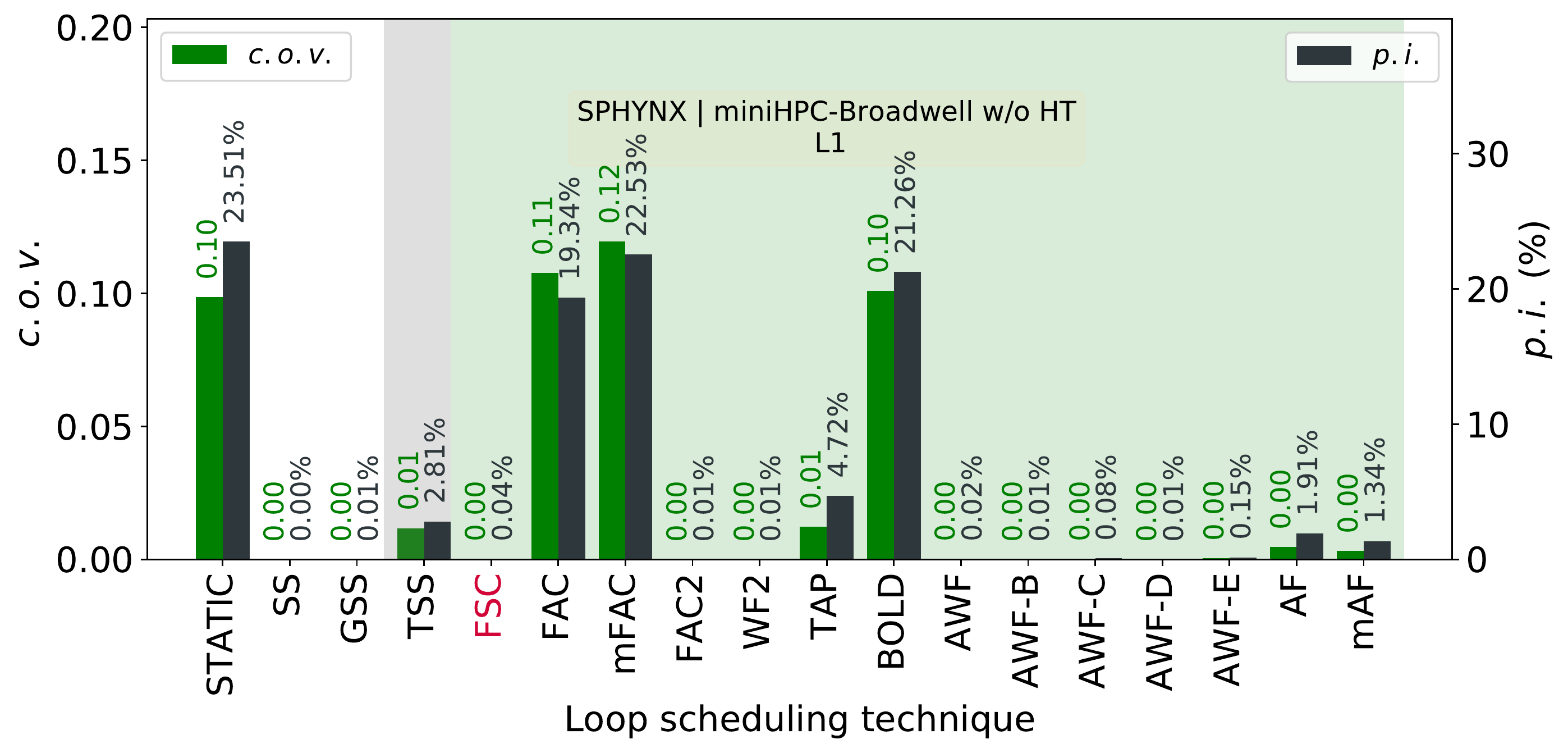}
	  			\label{fig:sphynxsmi}
	  		}

		
		\caption[]{
		Load imbalance metrics \cov and \percentimbalance for the most time consuming loop of SPHYNX from \figurename~\ref{fig:allresults} calculated based on the results obtained from \xeon without hyperthreading. 
		}
		\vspace{-0.4cm}
		\label{fig:lbmetrics}
\end{figure}

\figurename~\ref{fig:lbmetrics} presents the \textit{load imbalance metrics}, \cov and \percentimbalance, calculated for the most \mbox{time-consuming} loop of \sphynx execution on \xeon node.
These results show that most scheduling techniques achieve nearly perfect load balancing. 
\rwthree{
Although these applications are computationally-intensive, with few memory operations, we can observe that almost perfect load balancing does not directly translate to high performance due to the additional scheduling overhead and loss of data locality.
For instance, in \figurename~\ref{fig:lbmetrics}, \texttt{AWF-B} achieved perfect load balancing while in \figurename~\ref{fig:allresults} we can observe that the execution time of \sphynx executing on \xeon node with \texttt{AWF-B} was $21.40\%$ slower than \texttt{Best}.
}

\begin{figure}[!htb]
	\centering
	\subfigure{
		\includegraphics[width=0.7\linewidth]{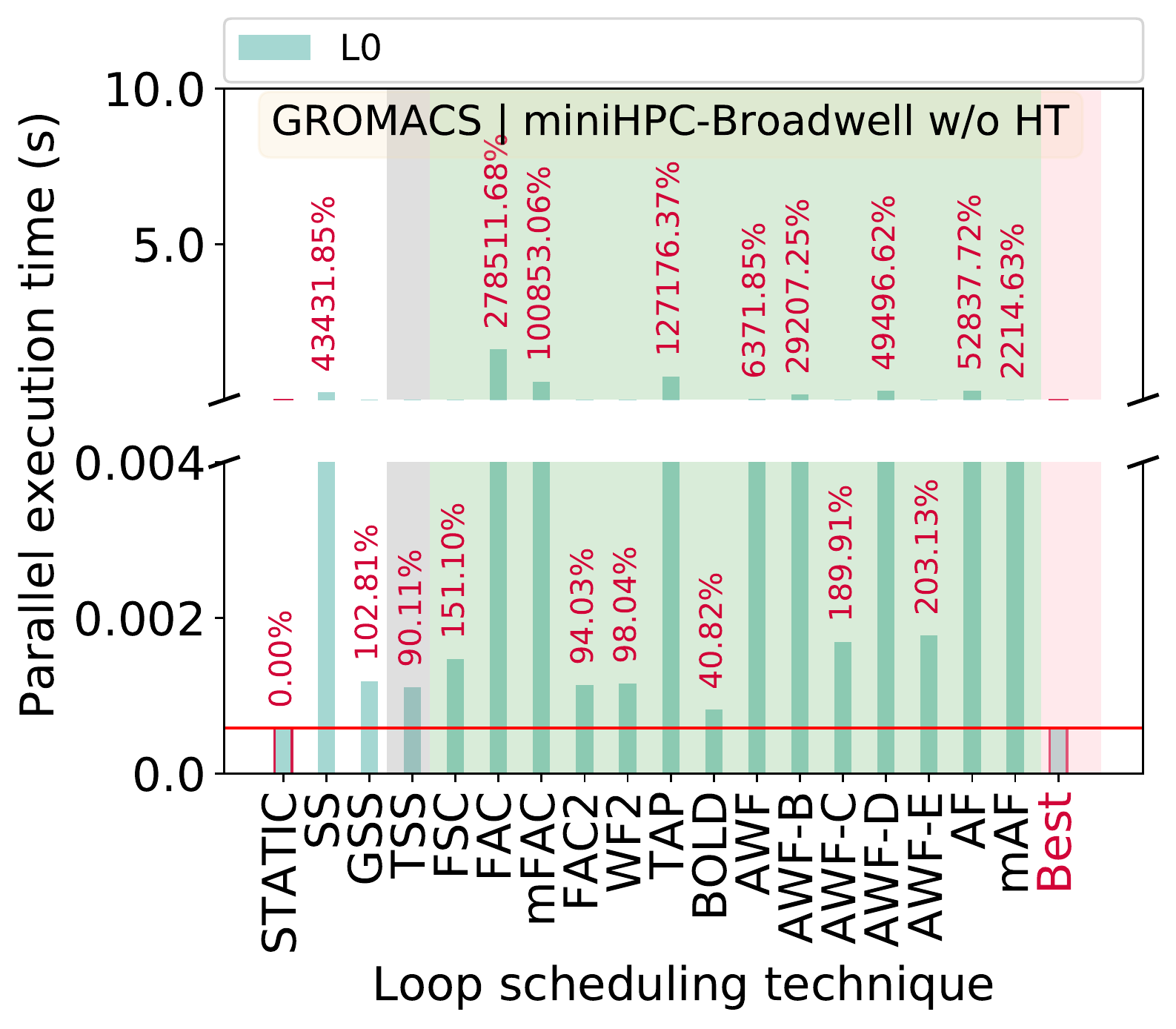}
		\label{fig:chunknab}
	}
	
	\caption[]{Parallel loop execution time for \gromacs' loop $L0$. The performance of the dynamic and adaptive self-scheduling techniques is compared to the performance of $STATIC$, which offers highest performance and data locality at lowest scheduling overhead. 
	}\vspace{-0.4cm}
	\label{fig:overhead}
\end{figure}

\rwthree{
Aspects such as scheduling overhead, ccNUMA effects, and data locality cannot directly be observed in~\figurename{~\ref{fig:allresults}}, as the scheduling overhead is absorbed by improvement in the loop execution time due to dynamic and adaptive scheduling. 
Instead, we use a computationally-inexpensive loop of a widely used molecular dynamics application \gromacs~\cite{gromacsPaper}, to reveal the overhead of all scheduling techniques. 
This particular loop in \gromacs has very low arithmetic intensity, regular loop iterations, and initializes three vector data structures which stresses ccNUMA effects and locality issues. 

We also use the \stream microbenchmark, a simple synthetic program that measures sustainable memory bandwidth, to show the memory bandwidth drop caused by ccNUMA effects and the locality issues that arise during dynamic and adaptive self-scheduling.

\figurename~\ref{fig:overhead} shows the parallel loop execution time for the loop from \gromacs executing on node \xeon while \figurename~\ref{fig:overheadstream} shows the memory bandwidth (in MB/s) maintained by each kernel from \stream also executing on node \xeon.
We use \xeon since it is a two-socket node, making ccNUMA effects and data locality issues more prominent.
}

Three factors contribute to the scheduling overhead shown in \figurename{~\ref{fig:overhead}}: 
(1)~Number of scheduling rounds~$o_{sr}$; 
(2)~Cost of calculating a chunk size~$o_{cs}$; and 
(3)~Synchronization cost between threads to obtain or to calculate a new chunk of loop iterations~$o_{sync}$.
Note that $o_{cs}$ and $o_{sync}$ are incurred with every scheduling round, therefore, growing with $o_{sr}$. 
\rwthree{\textbf{It is important to note that due to the non-deterministic and stochastic nature of dynamic and adaptive self-scheduling, high $o_{sr}$ potentiates the loss of data locality and the importance of ccNUMA effects, which in this case are compounded with the overhead (compared to \texttt{STATIC}) seen in \figurename{~\ref{fig:overhead}}.}}

\rwthree{\texttt{STATIC} in \figurename{~\ref{fig:overhead}} shows the \emph{smallest} scheduling overhead, where $o_{sr} = 1$, $o_{cs}$ is a simple and deterministic division, and $o_{sync} =0$, since threads need no synchronization. 

\texttt{SS} shows the largest overhead due to $o_{sr} \leq N$, $o_{cs}$ is negligible, and $o_{sync}$ can either be negligible if $o_{sr} \ll N$ or significant if $o_{sr} = N$. 
In specific situations, some techniques may show higher overhead than \texttt{SS} (e.g. in this case, \texttt{FAC}, \texttt{mFAC}, \texttt{TAP}, \texttt{AWF-D}, and \texttt{AF}). 
If a scheduling technique with higher $o_{cs}$ than \texttt{SS} calculates the same chunk size as \texttt{SS} it will cause more overhead than \texttt{SS}.

\texttt{GSS}, \texttt{TSS}, and \texttt{FSC} incur less overhead than \texttt{SS} due to assigning larger chunks of loop iterations, which reduces $o_{sr}<N$, increases data locality, while keeping $o_{cs}$ and $o_{sync}$ comparable by using simple chunk calculation functions and atomic operations to synchronize the threads.

}

\rwthree{The extremely large overhead with \texttt{FAC} ($>43,000\%$) is due to the combination of high $o_{sr}$, $o_{cs}$, $o_{sync}$ overheads and loss of data locality.
\texttt{FAC} uses a complex function to calculate the chunk size, requiring profiling information and a mutex to synchronize the threads.

\texttt{mFAC} has lower $o_{sync}$ than \texttt{FAC} by using atomic operations, leading to lower overall overhead than \texttt{FAC}. 
\texttt{FAC2} and \texttt{WF2} outperform \texttt{FAC} and \texttt{mFAC} in terms of $o_{cs}$ by using a simple chunk calculation function, not requiring profiling information, and using atomic operations for synchronization.

Similar to \texttt{FAC} and \texttt{mFAC}, \texttt{TAP} failed to calculate an appropriate chunk size based on the profiling information due to the very small loop iteration granularity, which resulted in loss of data locality, high $o_{sr}$, and $o_{cs}$.}


\rwthree{\texttt{BOLD} generates chunk sizes very similar to \texttt{STATIC} but at a very high chunk calculation cost, therefore, incurring high $o_{cs}$, and low $o_{sr}$ and $o_{sync}$.

\awfb and \awfd do not manage to adapt to the very fine iteration granularity of this \gromacs' loop and assign very small chunks, reducing data locality and increasing $o_{sr}$, and the cost of adaptation $o_{cs}$.
In contrast, \awfc and \awfe only incur from high $o_{cs}$.

\texttt{AF} and \texttt{mAF} also have very high $o_{cs}$. 
However, \texttt{mAF} also considers the time for $o_{cs}$ for the chunk size calculation.
Therefore, it increases its chunk size to reduce $o_{sr}$, offering improvement over \texttt{AF}.}

\rwthree{
In \figurename~\ref{fig:overhead} and \figurename~\ref{fig:overheadstream}, one can note that \texttt{SS} causes high scheduling overhead due to high $o_{sr}$ and loss of data locality, which justifies the low memory bandwidth shown in \figurename~\ref{fig:overheadstream} for all \stream kernels.
The low memory bandwidth achieved by \texttt{FAC} and \texttt{mFAC} is justifiable since those techniques not only cause high $o_{cs}$ and $o_{sync}$ overheads but also need to read profiling information collected on a separate execution of the application. 

\begin{figure}[!htb]
	\centering
	\subfigure{
		\includegraphics[width=0.98\linewidth]{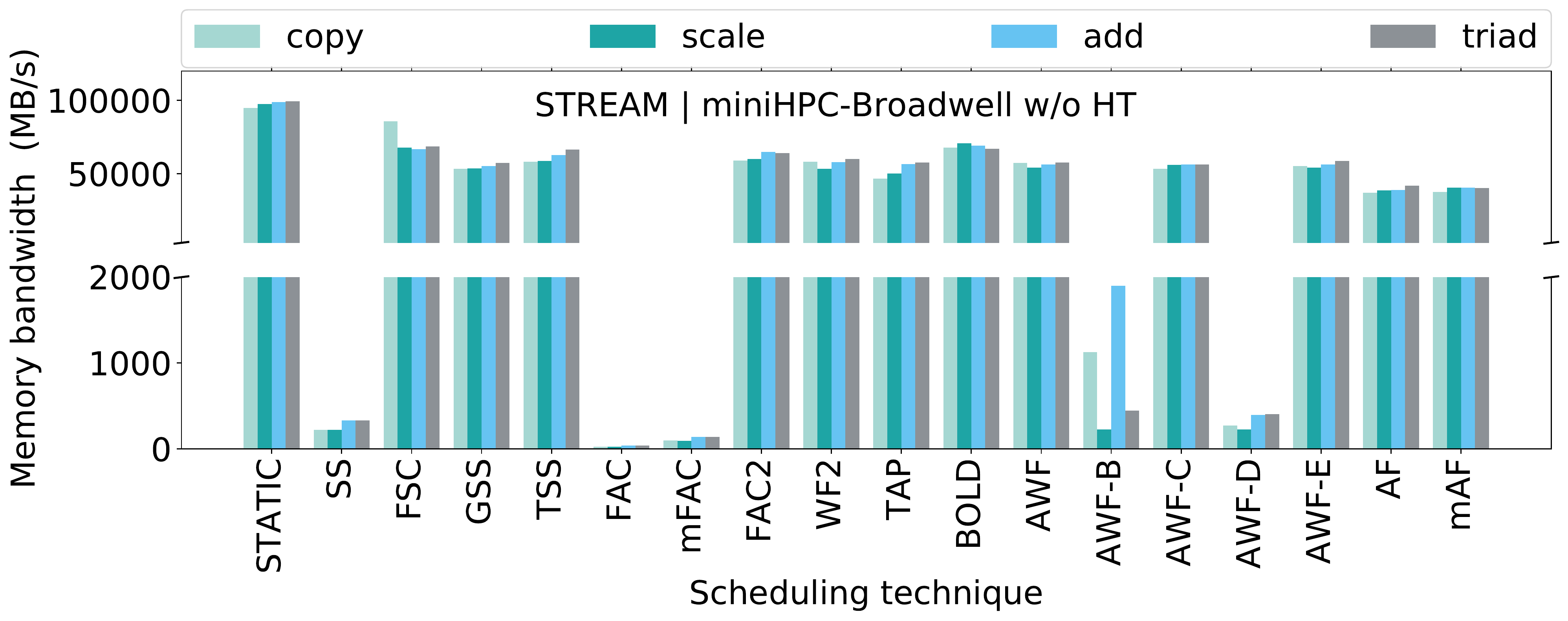}
		\label{fig:chunksphynx}
	}
	\caption[]{
	Memory bandwidth (MB/s) for executing STREAM with all scheduling techniques executing on \xeon.
	}
	\label{fig:overheadstream}
\end{figure}

For simple kernels, such as those from \stream and $L0$ from \gromacs, profiling the execution of each loop iteration may adversely influence execution performance, 
and may lead \texttt{FAC} and \texttt{mFAC} to calculate very small chunk sizes, increasing $o_{sr}$ and consequently also increasing scheduling overhead, ccNUMA effects, and loss of data locality.
This observation is also valid for dynamic and adaptive techniques which measure (during execution) the execution time of previous chunks of iterations to determine the next chunk size. 
If the loop kernel's arithmetic intensity is low, dynamic and adaptive techniques may measure inaccurate values which may result in small chunk sizes, higher $o_{sr}$, non-negligible ccNUMA effects and loss of data locality.
}

\vspace{-0.2cm}
\subsection{Impact of Chunk Parameter Choice}\label{sec:minchunk}

\begin{figure}[!htb]
	\centering
	
	
	\subfigure{
		\includegraphics[trim=0.2cm 0.6cm 0cm 0.25cm, clip=true, width=1\linewidth]{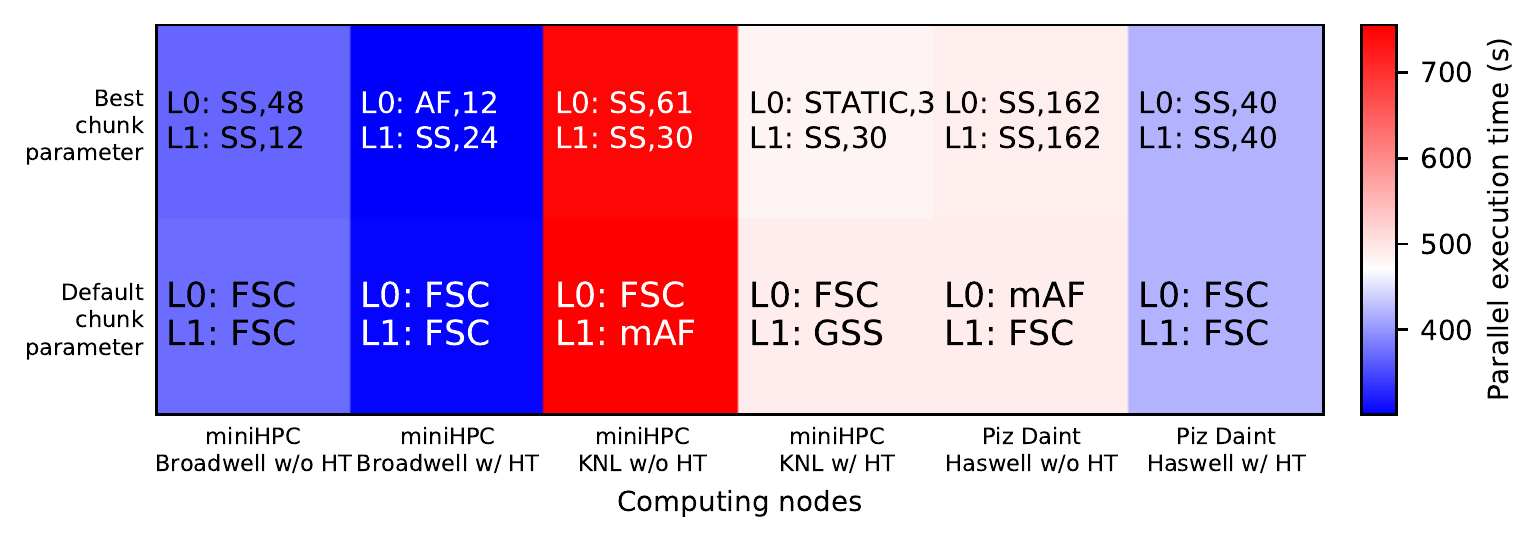}
		\label{fig:chunksphynx}
	}
	
	
	
	
	\caption[]{Comparison of the parallel execution time for the highest performing combination of DLS techniques with the default chunk parameter (\texttt{Best} in \figurename~\ref{fig:allresults}) vs. with the \emph{best} chunk parameter. 
	Shown is the parallel execution time (heat-bar) of \sphynx on various node types/configurations ($x$-axis) scheduled with the \texttt{Best} technique and two chunk parameters: default (chunk) vs. \emph{best} (chunk) ($y$-axis).
	}
	\label{fig:heatmapAllChunk}
\end{figure}

	We explore the performance impact of the chunk parameter for all DLS techniques. 
	We experimented with many different values for the chunk parameter per DLS technique, loop, application, and node type/configuration as indicated in Table~\ref{table:exp}. 
	\figurename~\ref{fig:heatmapAllChunk} presents an overview of the results for \sphynx comparing the \texttt{Best} combination of DLS techniques with the \emph{default} value of the chunk parameter vs. the most performing combination of DLS techniques with the \emph{best} value of the chunk parameter. 
	The \emph{best} value of the chunk parameter is identified by testing the application performance with values of the chunk parameter from $N/2P$ down to 1 (see Table~\ref{table:exp}).
 
 \begin{figure*}[!htb]
	\centering
	\adjustbox{width=1.2\textwidth}{
		\hspace{-2cm}		\begin{minipage}{1.3\textwidth}
	\subfigure[SPHYNX - $L1$ - \mbox{\xeon}]{
		\tcbox[colframe=highlight,
		colback=white, top=0pt,left=0pt,right=0pt,bottom=0pt]{\includegraphics[width=.31\linewidth]{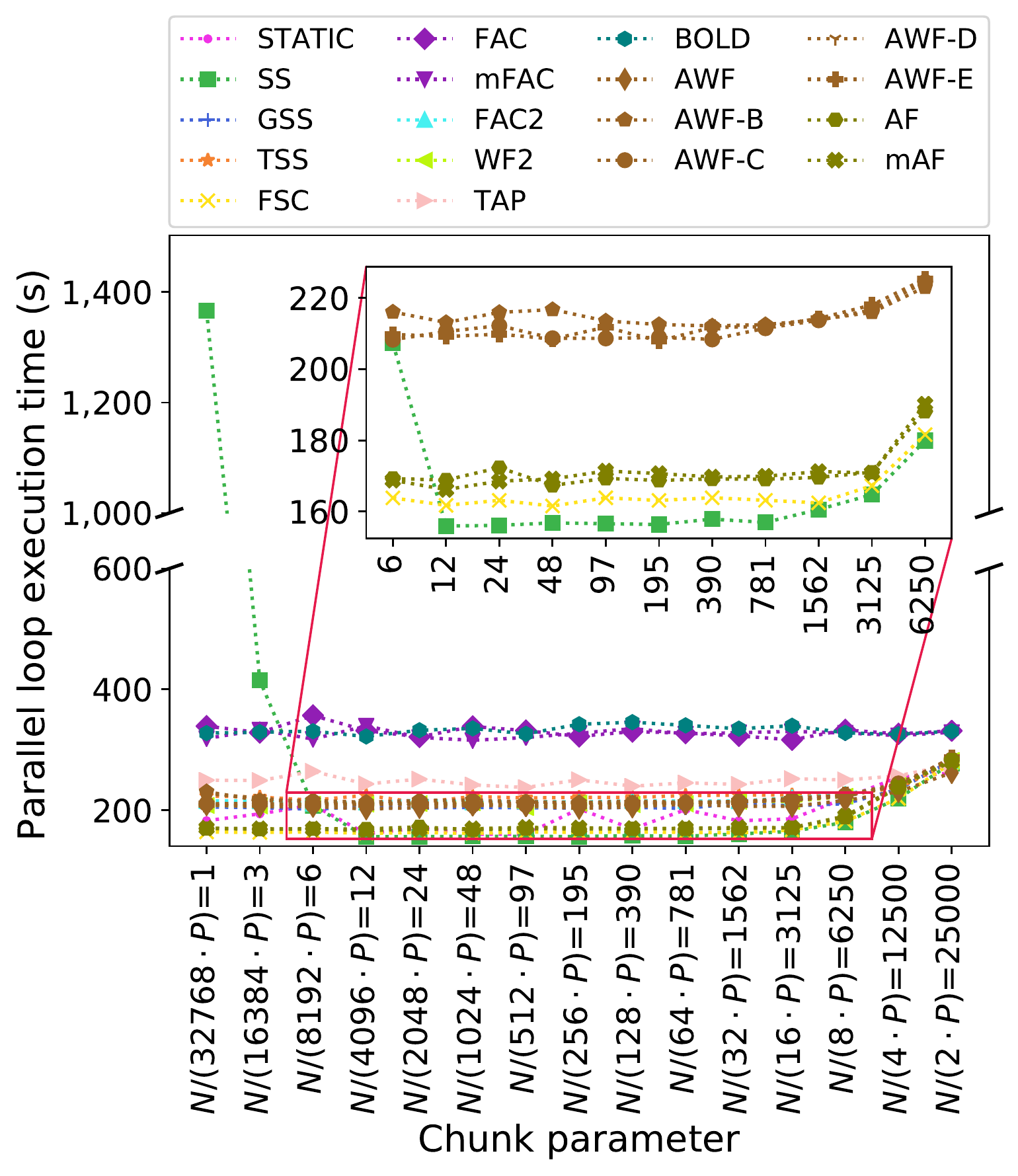}}
		\label{fig:chunkxeon}
	}
	\subfigure[SPHYNX - $L1$ - \mbox{\knl}]{
		\includegraphics[width=.31\linewidth]{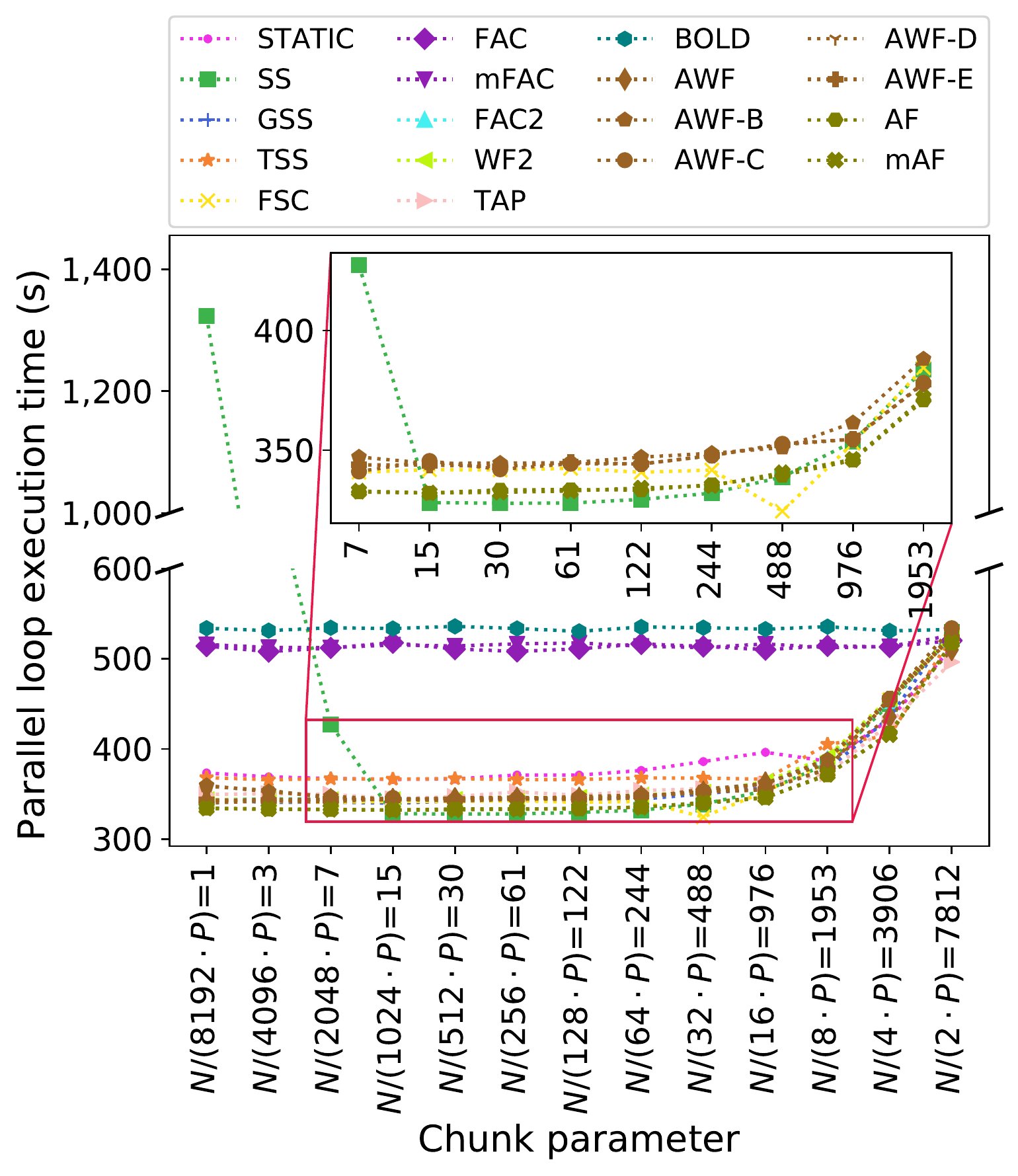}
		\label{fig:chunkknl}
	}
	\subfigure[SPHYNX - $L1$ - \mbox{\daint}]{
		\includegraphics[width=.31\linewidth]{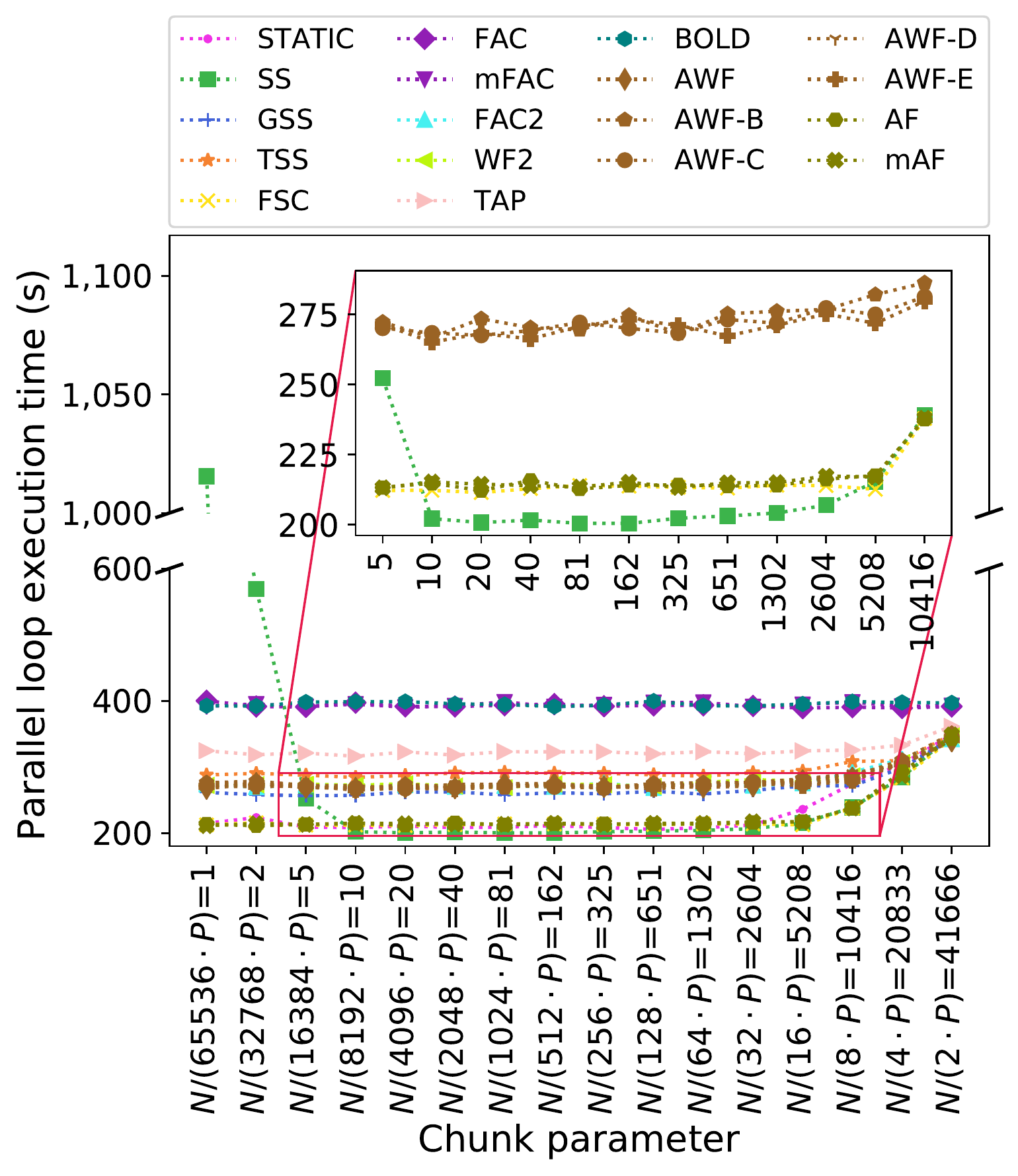}
		\label{fig:chunkpizdaint}
	}

	\caption[]{Parallel cumulative loop execution time for \textbf{\sphynx}'s $L1$ loop executing on \xeon, \knl, and \daint without hyperthreading. 
		The \textcolor{red}{red} rectangles are zoomed in for the range of chunk parameter values that achieved high performance to show the performance of \textit{dynamic} and \textit{adaptive} loop scheduling techniques \texttt{AWF-}\texttt{B},\texttt{C},\texttt{D},\texttt{E}, \texttt{AF}, and \texttt{mAF} and the \textit{dynamic} and \textit{non-\textit{adaptive}} loop scheduling techniques \texttt{SS} and \texttt{FSC}. 
		A proper chunk parameter value for \texttt{SS} reduces overall overhead and improves data locality, allowing \texttt{SS} to reach high performance or even to outperform all other techniques.}
	\end{minipage}
}
	\label{fig:chunksizes}
\end{figure*}
 
	In \figurename~\ref{fig:heatmapAllChunk}, the best chunk parameter always improved the performance of the applications.
	Similar colors indicate that the performance improvement was very low.
	A carefully selected chunk parameter for \texttt{SS} frequently achieves the highest performance. 
	However, the process of finding such an optimal value requires extensive experimentation (e.g., such as the experiments presented here), and must be performed for each loop and system that the application will execute on. 
	Furthermore, in the case of system variation, the optimal chunk parameter, once found, may no longer provide the highest performance since it would not be adapted during execution. 
	It is impractical to rely exclusively on a manual and extensive experimentation process to find an optimal chunk parameter. 
\rwthree{This makes \textbf{\textit{dynamically adaptive} loop scheduling techniques} a highly promising solution, especially on upcoming Exascale systems, which will increasingly be heterogeneous.}
	The existing dynamic and adaptive scheduling techniques offer a first step for performance \textit{auto-tuning} on a per loop basis against system and application variability.

	Based on the results in \figurename~\ref{fig:allresults} and \figurename~\ref{fig:heatmapAllChunk}, it is interesting to examine the impact of the chunk parameter on the performance of most time-consuming loop from \sphynx, $L1$.
	These results are shown in \figurename~\ref{fig:chunksizes},
	wherein the parallel execution time of $L1$ ($y$-axis) is shown for different chunk parameter values ($x$-axis).  
	The chunk parameter values differ for each system since they are calculated using the available number of threads (see Table~\ref{table:exp}). 
	
	
	In \figurename~\ref{fig:chunksizes} we expect to see \texttt{SS} reaching or outperforming \texttt{FSC}. 
	\rwthree{The \textit{dynamic} and \textit{adaptive} loop scheduling techniques are expected to improve the performance by reducing overhead since with a chunk parameter they preserve improved data locality, and are executed fewer times.}

	The performance of \texttt{SS} indeed reaches and outperforms that of \texttt{FSC} with larger chunk parameters between $12$ and $1,562$ for \xeon, $15$ and $122$ for \knl, and $10$ and $1,302$ for \daint. 
	This is due to the improved data locality and scheduling overhead of \texttt{SS} with a larger chunk parameter value.
	\texttt{FSC} is unaffected since it calculates a chunk size slightly larger than the range of chunk parameter values that achieve highest performance. 
	The performance of \texttt{FSC} is only affected when the chosen chunk parameter value is larger than the chunk size calculated by the technique itself. 
	
	All results in \figurename~\ref{fig:chunksizes} show that performance degrades with large chunk parameter values, approximately $2,000$ loop iterations. 
	This happens since the $L1$ loop from \mbox{\sphynx} is irregular (see \figurename~\ref{fig:lbmetrics}, $L1$ of \sphynx executed with \texttt{STATIC}) and, therefore, certain threads receive more work than others resulting in poor performance due to a load imbalanced execution. 
	We expected that the \textit{dynamic} and \textit{adaptive} loop scheduling techniques show improved performance when the chunk parameter is chosen since their overhead would be reduced \rwthree{while preserving data locality}. 
	This was not the case.
	These results are discussed in Section~\ref{sec:calcchunk}, where the progression of the chunk sizes of each DLS technique is explored, highlighting why no improvement can be observed for the \textit{dynamic} and \textit{adaptive} loop scheduling techniques in this particular case.

\subsection{Influence of Chunk Size Progression}\label{sec:calcchunk}

\begin{figure}[!htb]
	\centering
	\vspace{-0.2cm}\includegraphics[width=0.8\linewidth]{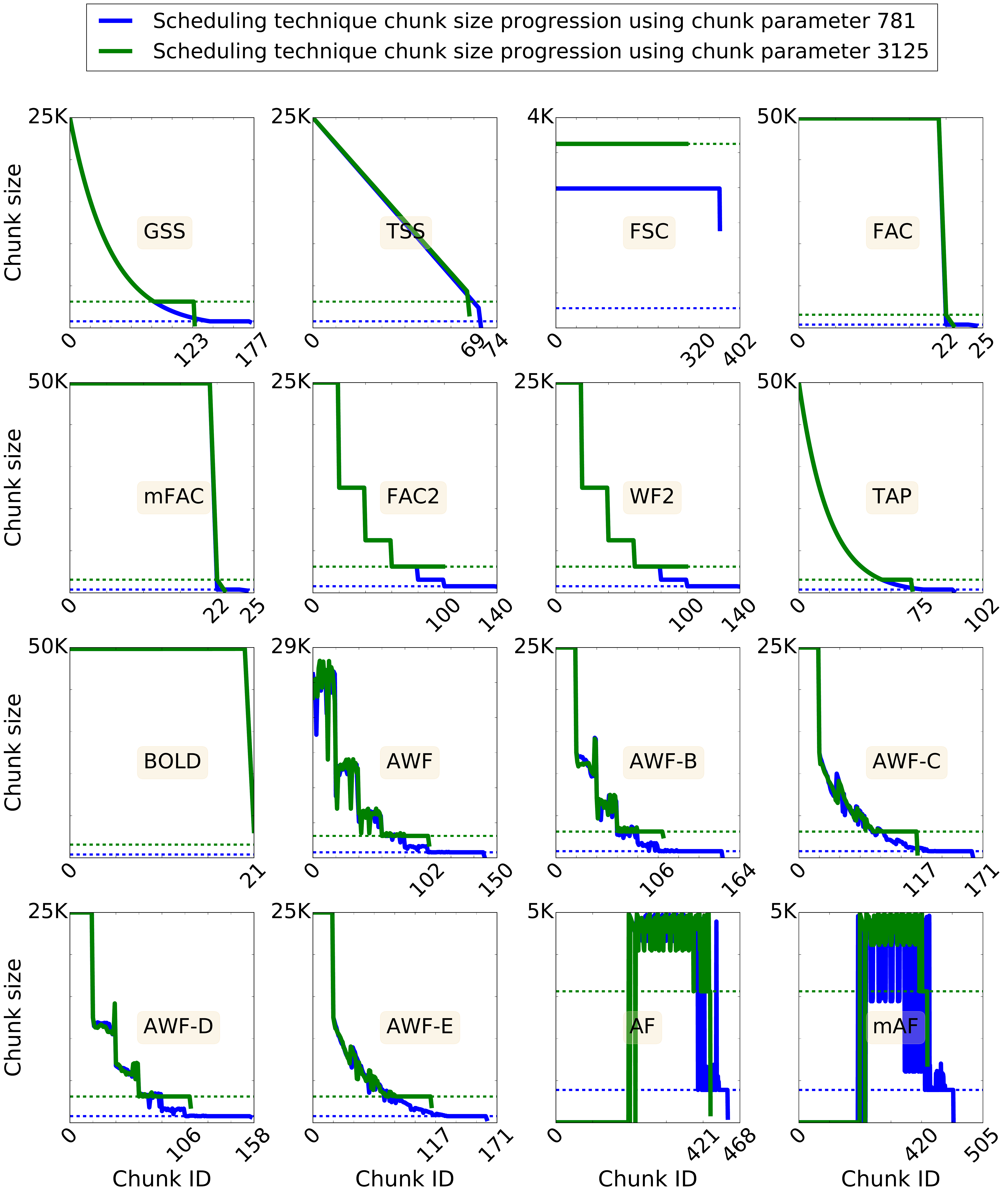}
	\caption{
		Progression of chunk sizes with the DLS techniques in \lbomp for scheduling \textbf{\sphynx}'s $L1$ loop of $1,000,000$ loop iterations on a 20-thread \xeon and chunk parameters $N/(64P) = 781$ iterations (\textcolor{blue}{blue} dotted line) and $N/(16P) = 3,125$ (\textcolor{darkgreen}{green} dotted line) iterations, respectively. 
		Chunk IDs are shown on the $x$ axis and their sizes on the $y$ axis.
		The \textcolor{blue}{blue} and \textcolor{darkgreen}{green} curves represent the progression of the chunk sizes produced by a DLS technique instantiated either with $781$ or $3,125$ as chunk parameter (lower threshold for calculating a chunk size). 
	}
	\vspace{-0.2cm}
	\label{fig:chunksize-id}
\end{figure}

The chunk size progression for the DLS techniques during the scheduling of the $L1$ loop from \sphynx is shown in \figurename~\ref{fig:chunksize-id}, with chunk ID on the $x$ axis (denoting the number of chunks produced) and their sizes on the $y$ axis. 
We used \lbomp with \texttt{KMP\_PRINT\_CHUNKS=1} (\textit{chunk information} feature, Section~\ref{sec:profile}) to collect and report the chunk sizes assigned by each DLS technique in every scheduling round.

	In general, fewer chunks imply \rwthree{improved data locality} and a smaller scheduling overhead due to fewer scheduling operations. 
	Fewer chunks may also result in potentially higher load imbalance since the chunk sizes are larger than with fewer chunks, as observed in \figurename~\ref{fig:chunksizes}, in Section~\ref{sec:minchunk}.
	The results for the \textit{dynamic} and \textit{adaptive} loop scheduling techniques \texttt{AWF-}\texttt{B},\texttt{C},\texttt{D},\texttt{E}, \texttt{AF}, and \texttt{mAF} in \figurename~\ref{fig:chunksize-id}, clarify why, in this case, no performance improvements can be observed when a chunk parameter is given. 
	Even with a relatively large value of the chunk parameter, such as $3,125$, none of the \textit{adaptive} loop scheduling techniques reaches the given value early enough to benefit from \rwthree{improved data locality} and the lower overhead of executing fewer scheduling rounds. 
	A much larger chunk parameter value would not necessarily improve loop performance due to the potential of load imbalance associated with large chunks.

	Apart from \texttt{AF}, \texttt{mAF}, and \texttt{FSC}, all scheduling techniques follow a decreasing chunk size pattern. 
	The \textit{dynamic} and \textit{adaptive} loop scheduling techniques \texttt{AWF-}\texttt{B},\texttt{C},\texttt{D},\texttt{E} follow a decreasing chunk size pattern (similar to \texttt{GSS} and \texttt{FAC2}), with the major difference of adapting to system variation by increasing or decreasing their chunk sizes \textit{during execution}.

    Each DLS technique produces a different number of chunks ($x$-axis) depending on its chunk calculation method (non-adaptive methods) and loop performance during execution (adaptive methods). 
    As described in Section~\ref{sec:perfeval}, the number of chunks is proportional to the number of scheduling rounds, $o_{sr}$, and contributes to the scheduling overhead associated with a technique.

	The first chunks calculated by \texttt{AF} and \texttt{mAF} are small as these DLS techniques perform a warm-up scheduling round where they gather initial information about the loop iterations performance. 
	These first chunks sizes are hard-coded to $10$ loop iterations and are unaffected by the declaration of the chunk parameter.


\vspace{-0.2cm}
\section{Conclusions and Future Work}\label{sec:conclusion}

We introduced \lbomp, a novel open-source library for dynamic load balancing of multithreaded applications that use \openmp, implemented as an extension of LLVM’s OpenMP RTL.
This work contributes: \emph{a systematic and unified implementation} of 14 \textit{dynamic} (and \textit{adaptive}) loop scheduling techniques; \emph{features for advanced performance measurement} of loop performance and load imbalance; and \emph{an \mbox{in-depth} analysis of the performance potential and limitations} of the \openmp standard and the newly implemented scheduling techniques. 
Through an extensive performance analysis campaign we showed that for numerous \mbox{application-systems} pairs, the scheduling techniques in \lbomp outperform those from the \openmp standard. 

\rwone{

With this work we \emph{bridge the gap} between the \mbox{state-of-the-art} and the \mbox{state-of-the-practice} of load balancing in multithreaded applications.
This will allow the efficient exploitation of large degrees of heterogeneous node-level parallelism for improving the performance of applications on upcoming Exascale systems.
}

\rwone{
\lbomp represents the \emph{first and necessary} step for devising automated methods to dynamically select the highest performing loop scheduling techniques during applications execution. 
Devising such methods is part of ongoing work by the authors.
}

\rwone{A possible extension is to expand the selection criteria to include additional DLS techniques in \lbomp.
The study of locality-aware self-scheduling techniques is a promising research direction.}
We plan to patch and up-stream the DLS techniques implemented in \lbomp to the main LLVM \openmp RTL, facilitating a broad use and impact for \openmp applications.
Applying \lbomp to \textit{explicit \openmp task scheduling} is also planned as future work.

\vspace{-0.2cm}
\section*{Acknowledgments}
This work has been in part supported by the Swiss National Science Foundation in the context of the ``Multi-level Scheduling in Large Scale High Performance Computers'' (MLS) grant, number 169123, the Swiss Platform for Advanced Scientific Computing (PASC) project ``SPH-EXA: Optimizing Smoothed Particle Hydrodynamics for Exascale Computing'', and by DAPHNE, funded by the European Union's Horizon 2020 research and innovation programme under grant agreement No 957407. We acknowledge access to Piz Daint at the Swiss National Supercomputing Centre, Switzerland under the PASC SPH-EXA's share with the project ID c16.
The authors also acknowledge Akan Yilmaz for his earlier contribution to this work.

\vspace{-0.25cm}

\bibliographystyle{IEEEtran}
\bibliography{llvm-dls-runtime}

\begin{thebibliography}{10}
\providecommand{\url}[1]{#1}
\csname url@samestyle\endcsname
\providecommand{\newblock}{\relax}
\providecommand{\bibinfo}[2]{#2}
\providecommand{\BIBentrySTDinterwordspacing}{\spaceskip=0pt\relax}
\providecommand{\BIBentryALTinterwordstretchfactor}{4}
\providecommand{\BIBentryALTinterwordspacing}{\spaceskip=\fontdimen2\font plus
\BIBentryALTinterwordstretchfactor\fontdimen3\font minus
  \fontdimen4\font\relax}
\providecommand{\BIBforeignlanguage}[2]{{%
\expandafter\ifx\csname l@#1\endcsname\relax
\typeout{** WARNING: IEEEtran.bst: No hyphenation pattern has been}%
\typeout{** loaded for the language `#1'. Using the pattern for}%
\typeout{** the default language instead.}%
\else
\language=\csname l@#1\endcsname
\fi
#2}}
\providecommand{\BIBdecl}{\relax}
\BIBdecl

\bibitem{hummel1996load}
S.~Flynn~Hummel, I.~Banicescu, C.-T. Wang, and J.~Wein, ``{Load Balancing and
  Data Locality via Fractiling: An Experimental Study},'' in \emph{{Lang.,
  Compi. RT. Sys. Scal. Compu.}}, 1996, pp. 85--98.

\bibitem{dongarra2011international}
J.~Dongarra, P.~Beckman, T.~Moore, P.~Aerts, G.~Aloisio, J.-C. Andre,
  D.~Barkai, J.-Y. Berthou, T.~Boku, B.~Braunschweig \emph{et~al.}, ``{The
  International Exascale Software Project Roadmap},'' \emph{{Intern. J. H. P.
  Comp. App.}}, pp. 3--60, 2011.

\bibitem{johnson1985np}
D.~S. Johnson, ``{The NP-completeness Column: An Ongoing Guide},'' \emph{{J. of
  Alg.}}, pp. 434--451, 1985.

\bibitem{khan1994comparison}
A.~Khan, C.~L. McCreary, and M.~S. Jones, ``{A Comparison of Multiprocessor
  Scheduling Heuristics},'' in \emph{{P. Intern. C. on Par. Proc.}}, vol.~2,
  1994, pp. 243--250.

\bibitem{izakian2009comparison}
H.~Izakian, A.~Abraham, and V.~Snasel, ``{Comparison of Heuristics for
  Scheduling Independent tasks on Heterogeneous Distributed Environments},'' in
  \emph{{Intern. C. Compu. Sci. Opt.}}, vol.~1, 2009, pp. 8--12.

\bibitem{bergman2008exascale}
K.~Bergman, S.~Borkar, D.~Campbell, W.~Carlson, and et~al., ``{Exascale
  Computing Study: Technology Challenges in Achieving Exascale Systems},''
  \emph{{Def. Adv. Research Proj. Ag. Info. Proc. Tech. Office, Tech. Rep}},
  2008.

\bibitem{asch2018big}
M.~Asch, T.~Moore, R.~Badia, M.~Beck, P.~Beckman, T.~Bidot, F.~Bodin,
  F.~Cappello, A.~Choudhary, B.~de~Supinski \emph{et~al.}, ``{Big data and
  Extreme-scale Computing: Pathways to Convergence-toward a Shaping Strategy
  for a Future Software and Data Ecosystem for Scientific Inquiry},''
  \emph{{Intern. J. H. P. Comp. App.}}, vol.~32, no.~4, pp. 435--479, 2018.

\bibitem{AF:2000}
I.~Banicescu and Z.~Liu, ``{Adaptive Factoring: A Dynamic Scheduling Method
  Tuned to the Rate of Weight Changes},'' in \emph{{P. of th H. P. C. Symp.}},
  2000, pp. 122--129.

\bibitem{Automatic-OMP-LS:2012}
P.~Thoman, H.~Jordan, S.~Pellegrini, and T.~Fahringer, ``{Automatic OpenMP Loop
  Scheduling: A Combined Compiler and Runtime Approach},'' in \emph{{P. Intern.
  W. on OpenMP}}, 2012, pp. 88--101.

\bibitem{Knowledge-BasedAdaptive2012}
Y.~Wang, W.~Ji, F.~Shi, Q.~Zuo, and N.~Deng, ``{Knowledge-Based Adaptive
  Self-Scheduling},'' in \emph{{P. Intern. C. on Net. and Par. Comp.}}, 2012,
  pp. 22--32.

\bibitem{ali2020twoLevel}
A.~Mohammed, A.~Cavelan, F.~M. Ciorba, R.~M. Cabez{\'o}n, and I.~Banicescu,
  ``{Two-level Dynamic Load Balancing for High Performance Scientific
  Applications},'' in \emph{{P. SIAM C. on Par. Proc. Sci. Comp.}}, 2020, pp.
  69--80.

\bibitem{Ayguade2003}
E.~Ayguad{\'e}, B.~Blainey, A.~Duran, J.~Labarta, F.~Mart{\'i}nez,
  X.~Martorell, and R.~Silvera, ``{Is the Schedule Clause Really Necessary in
  OpenMP?}'' in \emph{{P. Intern. W. on OpenMP App. and Tools}}, 2003, pp.
  147--159.

\bibitem{ciorba:2018}
F.~M. Ciorba, C.~Iwainsky, and P.~Buder, ``{OpenMP Loop Scheduling Revisited:
  Making a Case for More Schedules},'' in \emph{{P. Intern. W. on OpenMP}},
  2018.

\bibitem{Kasielke:2019}
F.~Kasielke, R.~Tsch\"uter, C.~Iwainsky, M.~Velten, F.~M. Ciorba, and
  I.~Banicescu, ``{Exploring Loop Scheduling Enhancements in OpenMP: An LLVM
  Case Study},'' in \emph{P. Intern. Symp. on Par. Dist. Comp.}, Amsterdam,
  2019.

\bibitem{heroux2020ecp}
M.~A. Heroux, J.~Carter, R.~Thakur, L.~McInnes, J.~Ahrens, T.~Munson,
  J.~Robert~Neely, and J.~S. Vetter, ``{ECP Software Technology Capability
  Assessment Report},'' {Oak Ridge National Lab.(ORNL), Oak Ridge, TN (United
  States)}, Tech. Rep., 2020.

\bibitem{FSC:1985}
C.~P. Kruskal and A.~Weiss, ``{Allocating Independent Subtasks on Parallel
  Processors},'' \emph{{J. Trans. on Soft. Eng.}}, pp. 1001--1016, 1985.

\bibitem{FAC:1992}
S.~Flynn~Hummel, E.~Schonberg, and L.~E. Flynn, ``{Factoring: A Method for
  Scheduling Parallel Loops},'' \emph{{J. of Comm.}}, pp. 90--101, 1992.

\bibitem{TAP:1992}
S.~Lucco, ``{A Dynamic Scheduling Method for Irregular Parallel Programs},'' in
  \emph{{P. C. on Progra. Lang. D. and Impl.}}, 1992, pp. 200--211.

\bibitem{WF:1996}
S.~Flynn~Hummel, J.~Schmidt, R.~N. Uma, and J.~Wein, ``{Load-sharing in
  Heterogeneous Systems via Weighted Factoring},'' in \emph{{P. Symp. on Par.
  Alg. Arch.}}, 1996, pp. 318--328.

\bibitem{BOLD:1997}
T.~Hagerup, ``{Allocating Independent Tasks to Parallel Processors: An
  Experimental Study},'' \emph{{J. Par. and Dist. Comp.}}, pp. 185--197, 1997.

\bibitem{AWF:2003}
I.~Banicescu, V.~Velusamy, and J.~Devaprasad, ``{On the Scalability of Dynamic
  Scheduling Scientific Applications with Adaptive Weighted Factoring},''
  \emph{{J. of Clus. Comp.}}, pp. 215--226, 2003.

\bibitem{HSS2006}
A.~Kejariwal, A.~Nicolau, and C.~D. Polychronopoulos, ``History-aware
  self-scheduling,'' in \emph{P. Intern. C. on Par. Proc.}, 2006, pp. 185--192.

\bibitem{penna2017binlpt}
P.~H. Penna, M.~Castro, P.~Plentz, H.~C. Freitas, F.~Broquedis, and J.-F.
  M{\'e}haut, ``{BinLPT: A Novel Workload-aware Loop Scheduler for Irregular
  Parallel Loops},'' \emph{{Simp. em Sis. Comp. de Alto Desemp.}}, 2017.

\bibitem{facProb2020}
K.-R. Kim, K.~Youngjae, and S.~Park, ``A probabilistic machine learning
  approach to scheduling parallel loops with bayesian optimization,'' \emph{J.
  Trans. on Par. and Dist. Sys.}, 2020.

\bibitem{SeonmyeongOPTexecParUDS}
S.~Bak, Y.~Guo, P.~Balaji, and V.~Sarkar, ``{Optimized Execution of Parallel
  Loops via User-Defined Scheduling Policies},'' in \emph{{P. Intern. C. on
  Par. Proc.}}, 2019.

\bibitem{kale2019toward}
V.~Kale, C.~Iwainsky, M.~Klemm, J.~H. Kornd\"orfer~M\"uller, and F.~M. Ciorba,
  ``{Towards A Standard Interface for User-Defined Scheduling in OpenMP},'' in
  \emph{{P. Intern. W. on OpenMP}}, 2019.

\bibitem{santana:hal-02454426}
\BIBentryALTinterwordspacing
A.~Santana, V.~Freitas, M.~Castro, L.~Lima~Pilla, and J.-F. M{\'e}haut,
  ``{ARTful: A Specification for User-defined Schedulers Targeting Multiple HPC
  Runtime Systems},'' 2020. [Online]. Available:
  \url{https://hal.archives-ouvertes.fr/hal-02454426}
\BIBentrySTDinterwordspacing

\bibitem{TSS:1993}
T.~H. Tzen and L.~M. Ni, ``{Trapezoid Self-scheduling: A Practical Scheduling
  Scheme for Parallel Compilers},'' \emph{J. Trans. on Par. Dist. Sys.}, pp.
  87--98, 1993.

\bibitem{FSC-DataLoc:1996}
M.~D. Durand, T.~Montaut, L.~Kervella, and W.~Jalby, ``{Impact of Memory
  Contention on Dynamic Scheduling on NUMA Multiprocessors},'' in \emph{P.
  Intern. C. on Par. Proc.}, 1993, pp. 258--262.

\bibitem{SS}
T.~Peiyi and Y.~Pen-Chung, ``{Processor Self-Scheduling for Multiple-Nested
  Parallel Loops},'' in \emph{{P. Intern. C. on Par. Proc.}}, 1986, pp.
  528--535.

\bibitem{GSS:1987}
C.~D. Polychronopoulos and D.~J. Kuck, ``{Guided Self-Scheduling: A Practical
  Scheduling Scheme for Parallel Supercomputers},'' \emph{J. Trans. on Compu.},
  pp. 1425--1439, 1987.

\bibitem{duran2008evaluation}
A.~Duran, J.~Corbal{\'a}n, and E.~Ayguad{\'e}, ``{Evaluation of OpenMP Task
  Scheduling Strategies},'' in \emph{{Intern. W. on OpenMP}}, 2008, pp.
  100--110.

\bibitem{clet2014evaluation}
J.~Clet-Ortega, P.~Carribault, and M.~P{\'e}rache, ``{Evaluation of OpenMP Task
  Scheduling Algorithms for Large NUMA Architectures},'' in \emph{{Eu. C. on
  Par. Proc.}}, 2014, pp. 596--607.

\bibitem{eleliemy2021distributed}
A.~Eleliemy and F.~M. Ciorba, ``{A Distributed Chunk Calculation Approach for
  Self-scheduling of Parallel Applications on Distributed-memory Systems},''
  \emph{{J. of Computa. Sci.}}, p. 101284, 2021.

\bibitem{derose2007detecting}
L.~DeRose, B.~Homer, and D.~Johnson, ``{Detecting Application Load Imbalance on
  High End Massively Parallel Systems},'' in \emph{Eu. C. on Par. Proc.}\hskip
  1em plus 0.5em minus 0.4em\relax Springer, 2007, pp. 150--159.

\bibitem{gromacsPaper}
D.~Van Der~Spoel, E.~Lindahl, B.~Hess, G.~Groenhof, A.~E. Mark, and H.~J.
  Berendsen, ``{GROMACS:} fast, flexible, and free,'' \emph{J. of compu.
  chemistry}, pp. 1701--1718, 2005.

\end{thebibliography}

%



\begin{wrapfigure}{l}{115mm} 
	\includegraphics[width=1in,height=1in,clip,keepaspectratio]{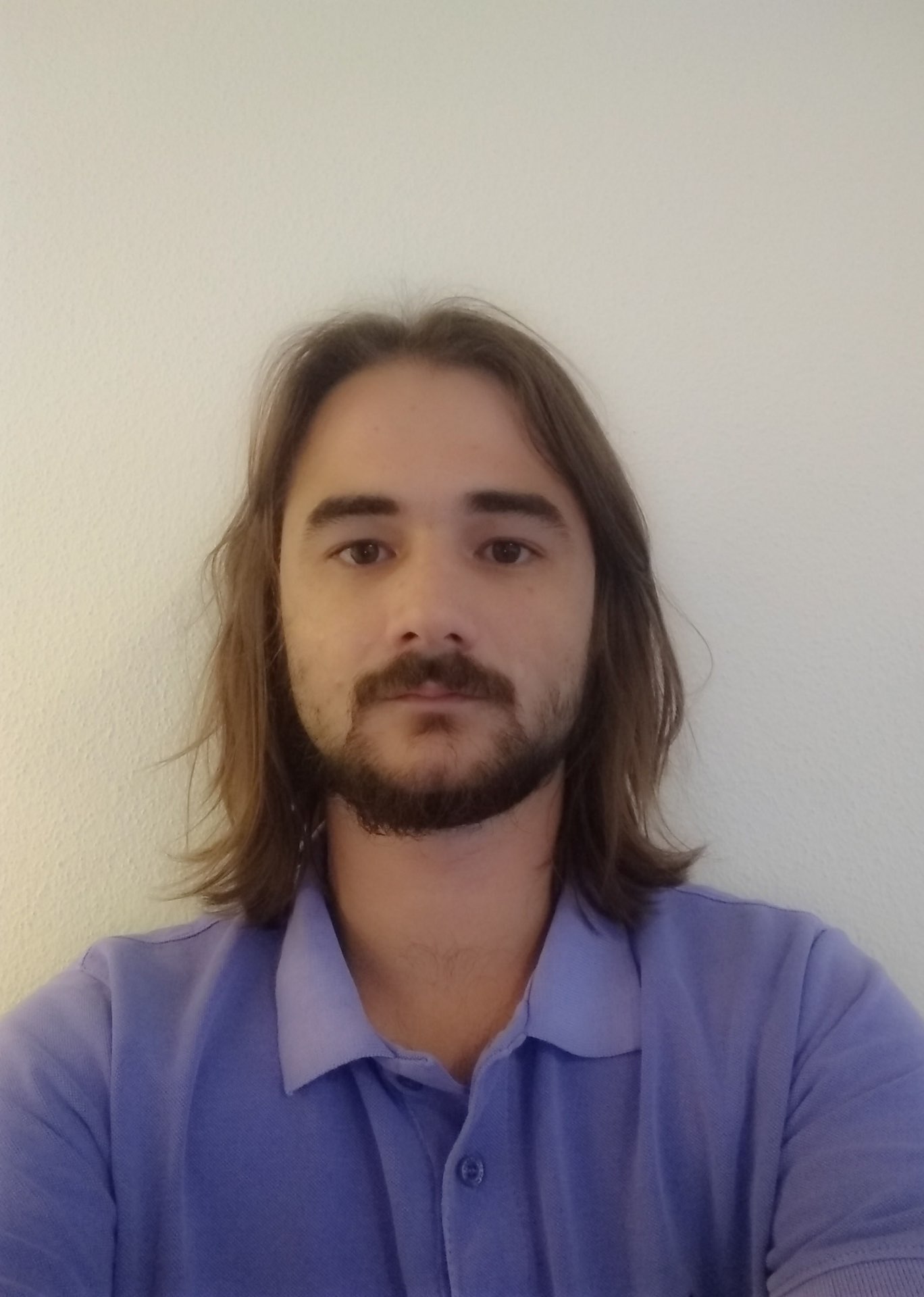}{Jonas H. M{\"u}ller Kornd{\"o}rfer} is a PhD candidate at the Department of Mathematics and Computer Science at the University of Basel, Switzerland. 
His main research interests include load balancing, scheduling, and mapping of computation and communication intensive applications. Website: \href{https://hpc.dmi.unibas.ch/en/people/jonas-h-mueller-korndoerfer}{hpc.dmi.unibas.ch/en/people/jonas-h-mueller-korndoerfer}.%
\end{wrapfigure}\par

\begin{wrapfigure}{l}{115mm} 
\includegraphics[width=1in,height=1in,clip,keepaspectratio]{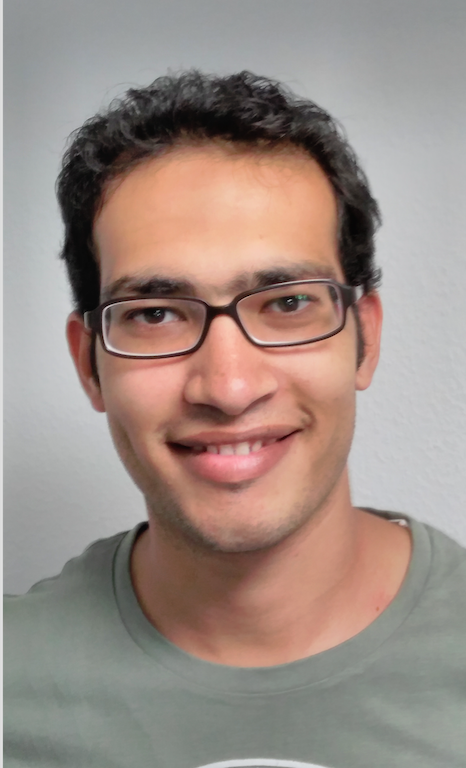}{Ahmed Eleliemy} is a postdoctoral researcher at the High Performance Computing Group at the Department of Mathematics and Computer Science at the University of Basel, Switzerland. He received his doctoral degree in multilevel scheduling of computations on large-scale parallel systems from the University of Basel in 2021. Website: \href{https://hpc.dmi.unibas.ch/en/people/ahmed-eleliemy}{hpc.dmi.unibas.ch/en/people/ahmed-eleliemy}.
\end{wrapfigure}\par

\begin{wrapfigure}{l}{115mm} 
	\includegraphics[width=1in,height=1in,clip,keepaspectratio]{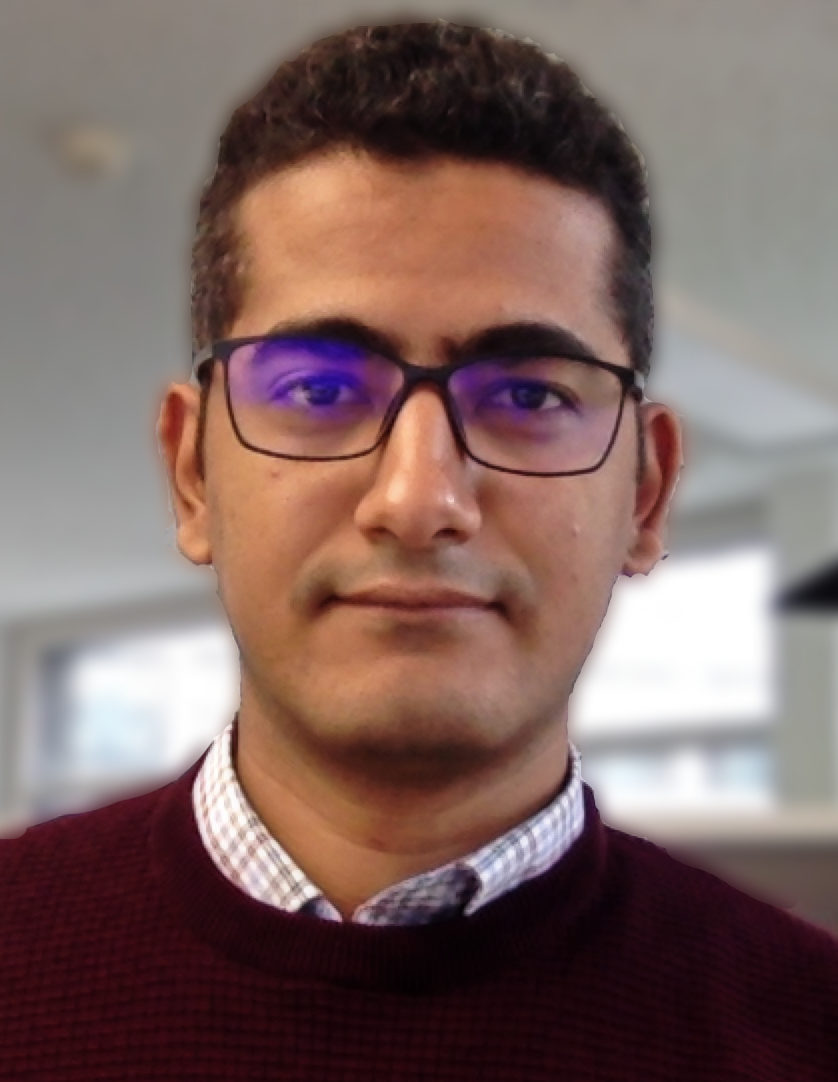}{Ali Mohammed} is a research engineer at HPE’s HPC/AI EMEA Research Lab (ERL), Switzerland. From March 2020 to April 2021, he was a postdoctoral researcher at the High-Performance Computing group at the University of Basel, Switzerland. He received his doctoral degree in robust scheduling for high performance computing from University of Basel in 2020. 
\end{wrapfigure}\par

\begin{wrapfigure}{l}{115mm} 
	\includegraphics[width=1in,height=1in,clip,keepaspectratio]{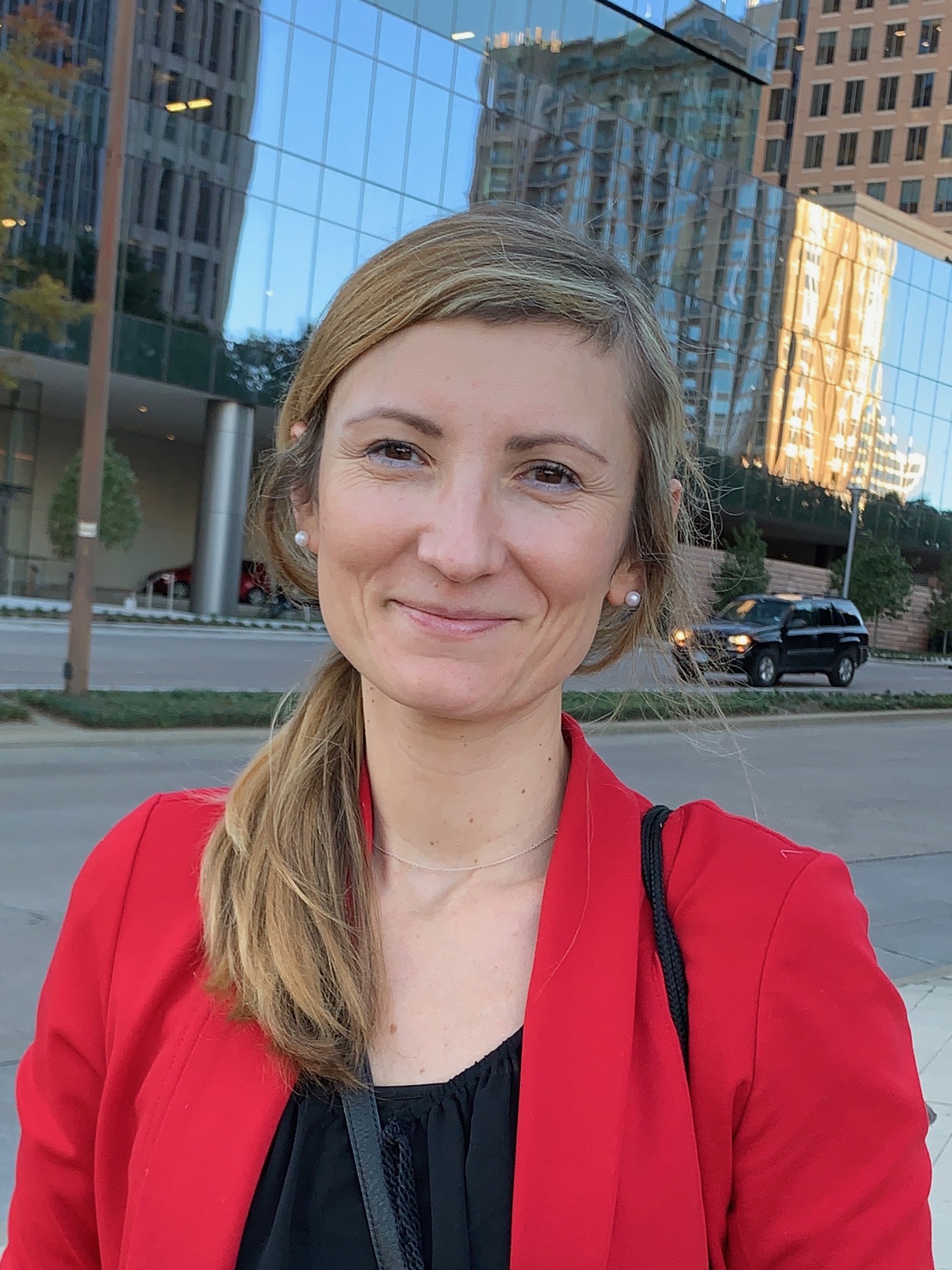}{Florina M. Ciorba} is an Associate Professor of High Performance Computing at the University of Basel, Switzerland. 
Her research interests include exploiting multilevel/hierarchical parallelism, dynamic and adaptive load balancing and scheduling, robustness, resilience, scalability, reproducibility, and benchmarking. 
Website: \href{https://hpc.dmi.unibas.ch/en/people/florina-ciorba}{hpc.dmi.unibas.ch/en/people/florina-ciorba}.
\end{wrapfigure}\par

\end{document}